\documentclass[%
 aip,
 amsmath,amssymb,
 reprint,%
]{revtex4-2}

\usepackage{float}
\usepackage{multibib}
\usepackage{hyperref}
\usepackage{xcolor}
\hypersetup{
    colorlinks,
    linkcolor={blue!50!blue},
    citecolor={blue!50!blue},
    urlcolor={blue!80!black}
}
\usepackage{siunitx}
\usepackage{import}
\usepackage{placeins}
\usepackage{xcolor}
\usepackage{lmodern}
\usepackage{graphicx}% Include figure files
\usepackage{dcolumn}% Align table columns on decimal point
\usepackage{bm}% bold math

\begin{document}

%\preprint{APS/123-QED}

\title{Localized creation of yellow single photon emitting carbon complexes in hexagonal boron nitride}

\author{Anand Kumar}
\email{anand.kumar@uni-jena.de}
\author{Chanaprom Cholsuk}
\author{Ashkan Zand}
\author{Mohammad N. Mishuk}
\author{Tjorben Matthes}
\affiliation{Abbe Center of Photonics, Institute of Applied Physics, Friedrich Schiller University Jena, 07745 Jena, Germany}
\author{Falk Eilenberger}
\affiliation{Abbe Center of Photonics, Institute of Applied Physics,
Friedrich Schiller University Jena, 07745 Jena, Germany}
\affiliation{Fraunhofer Institute for Applied Optics and Precision Engineering, 07745 Jena, Germany}
\affiliation{Max Planck School of Photonics, 07745 Jena, Germany}

\author{Sujin Suwanna}
\affiliation{Optical and Quantum Physics Laboratory, Department of Physics, Faculty of Science, Mahidol University, Bangkok 10400, Thailand}

\author{Tobias Vogl}%
\email{tobias.vogl@uni-jena.de}
\affiliation{Abbe Center of Photonics, Institute of Applied Physics,
Friedrich Schiller University Jena, 07745 Jena, Germany}
\affiliation{Fraunhofer Institute for Applied Optics and Precision Engineering, 07745 Jena, Germany}

\date{\today}% It is always \today, today,

\begin{abstract}
Single photon emitters in solid-state crystals have received a lot of attention as building blocks for numerous quantum technology applications. Fluorescent defects in hexagonal boron nitride (hBN) stand out due to their high luminosity and robust operation at room temperature. The fabrication of identical emitters at pre-defined sites is still challenging, which hampers the integration of these defects in optical systems and electro-optical devices. Here, we demonstrate the localized fabrication of hBN emitter arrays by electron beam irradiation using a standard scanning electron microscope with deep sub-micron lateral precision. The emitters are created with a high yield and a reproducible spectrum peaking at 575 nm. Our measurements of optically detected magnetic resonance have not revealed any addressable spin states. Using density functional theory, we attribute the experimentally observed emission lines to carbon-related defects, which are activated by the electron beam. Our scalable approach provides a promising pathway for fabricating room temperature single photon emitters in integrated quantum devices.
\end{abstract}

\keywords{Scanning electron microscope, density functional theory, carbon-based defects, localized fabrication, optically detected magnetic resonance.}

\maketitle

%\tableofcontents

\section{Introduction}

\begin{figure*}
    \includegraphics[width = 1\textwidth]{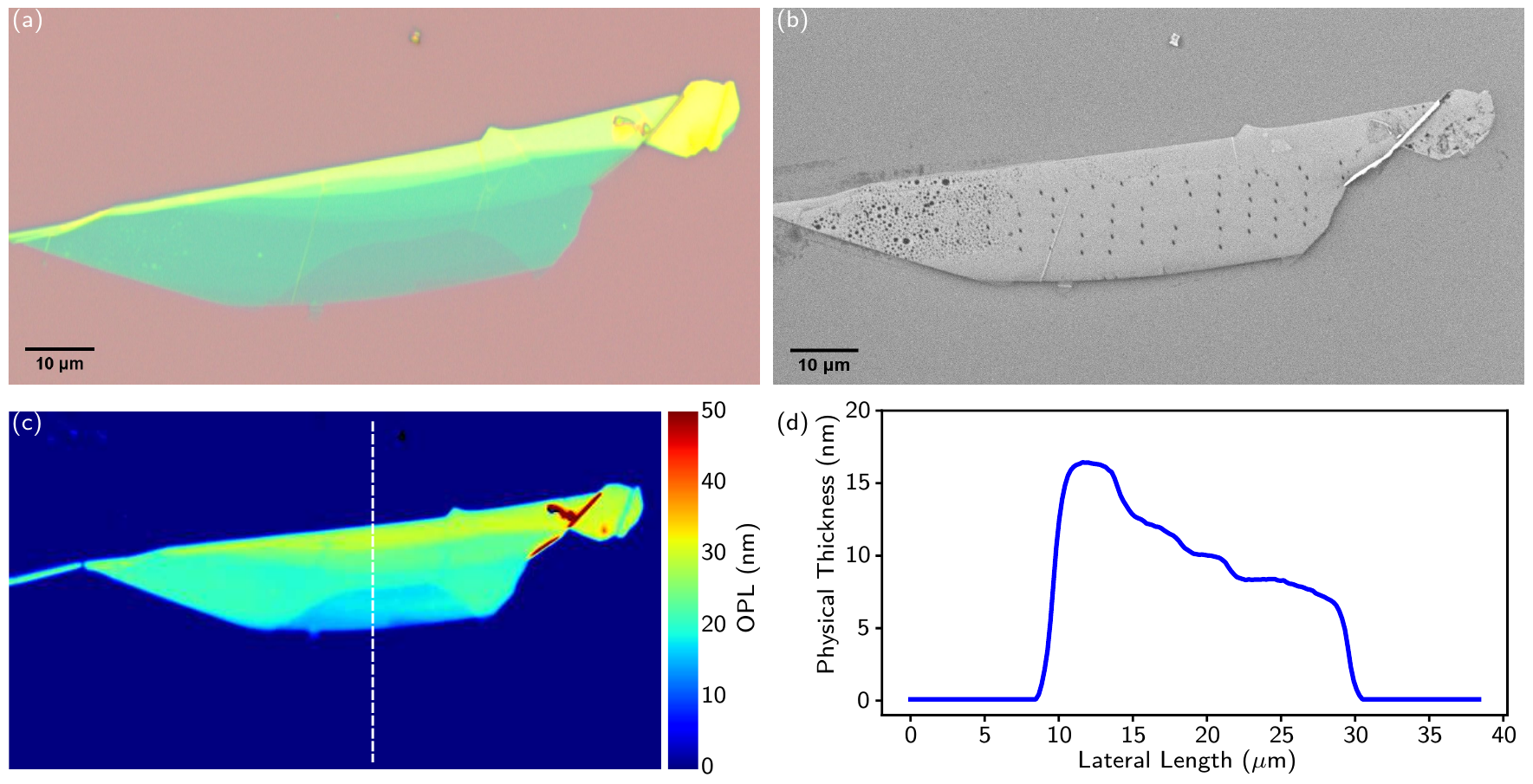}
    \caption{(a) Optical microscope image of the exfoliated hBN flake on a Si/SiO$_2$ substrate. (b) SEM image of the flake after electron irradiation. The radiation sites are visible as distinctive dots. The image was created with a low electron dose to avoid the random creation of emitters. (c) Height map of the hBN flake with the color-bar depicting the optical path length (OPL) measured with a PSI. (d) Physical thickness of the flake calculated along the dashed line using RCWA simulations.}
    \label{fig-1}
\end{figure*}
Many quantum photonic technologies, such as optical quantum computing and quantum communication, require sources of true single photons \cite{obrien_furusawa_vuckovic_2009, PhysRevResearch.2.023378, aspuru-guzik_walther_2012, cai_retzker_jelezko_plenio_2013, lvovsky_sanders_tittel_2009, Lo_2014}. The generation of ideal single photons on demand, however, still remains a technical challenge. Due to their in principle simple operation requirements and possibilities for integration, solid-state color centers are promising candidates for single photon sources as opposed to semiconductor quantum dots. While the latter can emit simultaneously pure and indistinguishable single photons \cite{10.1038/nnano.2017.218}, the need for cryogenic cooling below 4 K prevents using quantum dots in broad applications. Within the class of solid-state single photon emitters (SPEs) \cite{aharonovich_englund_toth_2016}, color centers hosted by two-dimensional (2D) materials have the advantage of featuring an in-plane dipole that emits out-of-plane (in case of intra-layer defects) \cite{C9NR04269E,doi:10.1126/sciadv.aba6038,doi:10.1063/5.0074946}. This simplifies the extraction and collection of the generated single photons, together with the fact that in atomically thin materials the emitters are not surrounded by any high refractive index material and are thus not affected by total internal or Fresnel reflection.\\
 \indent Single photon emitters hosted by 2D crystals have been observed initially in semiconducting transition metal dichalcogenides (TMDs) \cite{klein_lorke_florian_sigger_sigl_rey_wierzbowski_cerne_muller_mitterreiter_et, klein_sigl_gyger_barthelmi_florian_rey_taniguchi_watanabe_jahnke_kastl_et,srivastava_sidler_allain_lembke_kis_imamoglu_2015, he_clark_schaibley_he_chen_wei_ding_zhang_yao_xu_et,Tonndorf15, Carmen2017, Branny2017} and in insulating hexagonal boron nitride (hBN) \cite{tran_bray_ford_toth_aharonovich_2016}. The wide bandgap of hBN allows for the formation of deep defect-induced energy levels that are well-isolated from the band edges. This in turn leads to a high quantum efficiency at room temperature \cite{Nikolay:19} and is one of the reasons for the high single photon luminosity. Also contributing to this is a short excited-state lifetime \cite{vogl_campbell_buchler_lu_lam_2018} and the aforementioned high extraction efficiency. Unlike the nitrogen vacancy center in diamond, hBN emitters in the visible spectrum often have a strong coupling into the zero-phonon line (ZPL) with only weak phonon coupling that leads to coherent emission \cite{kubanek_2022}.\\
\indent The quantum emitters in hBN can be fabricated in a rich variety of ways: thermal activation of naturally occurring emitters \cite{tran_bray_ford_toth_aharonovich_2016}, plasma and chemical etching \cite{vogl_campbell_buchler_lu_lam_2018,10.1021/acs.nanolett.6b03268}, nanoindentation with an atomic force microscope (AFM) \cite{xu_martin_sychev_lagutchev_chen_taniguchi_watanabe_shalaev_boltasseva_2021}, strain-activation through transfer \cite{10.1364/OPTICA.5.001128} or growth on pillared substrates \cite{doi:10.1021/acs.nanolett.1c00685}, laser or gamma-ray damage \cite{Hou_2017,vogl_sripathy_sharma_reddy_sullivan_machacek_zhang_karouta_buchler_doherty}, and irradiation with electrons or ions \cite{10.1021/acsami.6b09875,naturecomm, doi:10.1021/acsphotonics.2c00631, PhysRevB.100.155419, ngoc}. While the strain activation localizes the emitter by default, many of the other fabrication techniques cause the emitter to form at random locations. Radiation damage has the potential for localization as well, since a laser or a charged particle beam can be easily focused onto a spot. The diffraction limit of electrons and ions is dependent on the particle energy and is typically on the order of a few nm, which may allow one to fabricate emitters with extremely high accuracy.\\
\indent In order to build integrated quantum devices with a scalable fabrication process, this precise control over the emitter location is crucial. Moreover, depending on the application, spectral control is also important. Quantum emitters have been experimentally reported with emission wavelengths in the UV \cite{doi:10.1021/acs.nanolett.6b01368}, the entire visible spectrum \cite{tran_elbadawi_totonjian_lobo_grosso_moon_englund_ford_aharonovich_toth_et}, and in the NIR \cite{doi:10.1063/5.0008242}. Potentially, this can be extended to telecom wavelengths \cite{Shaik2022}. In addition, theoretical simulations have shown that hBN emitters cover important wavelengths for quantum technologies \cite{Cholsuk2022}. Spectral control is difficult to achieve because the nature and type of these defects is still a topic of ongoing research. Recent studies supported by density functional theory (DFT) calculations have identified carbon- and oxygen-related defects, as well as the negatively charged boron vacancy \cite{Mendelson2021,doi:10.1126/sciadv.abe7138,Ivady2020,Auburger2021,Mackoit2019}.\\
\indent In this work, we address both issues of position and spectral control in the fabrication of quantum emitters in hBN. Using localized electron irradiation in a scanning electron microscope (SEM), we realize localized quantum emitters with a high yield. The high lateral resolution of the SEM allows us to restrict the emitter formation with sub-micron precision at pre-defined spots. While previous studies have found this to work reliably for blue emitters \cite{naturecomm,doi:10.1021/acsphotonics.2c00631}, our method expands this to yellow emitters with a reproducible wavelength and also does not require high temperature annealing (similar to other emitters fabricated by electron irradiation). We have also investigated the defects for addressable spin states through optically detected magnetic resonance (ODMR) \cite{gottscholl,stern, liu_feng_wang_li_liu_2019}. Together with DFT calculations, we propose a possible atomic structure of the defect complex with a ZPL close to the yellow emitters.

\section{Methods}
An hBN flake is mechanically exfoliated from a bulk crystal using scotch tape onto a PDMS (polydimethylsiloxane) stamp. Thin flakes (typically in the range of 5 nm to 20 nm) are selected based on the optical contrast in a bright field microscope and transferred to a grid-patterned Si/SiO$_2$ substrate (see Supplementary Section S1). Figures \ref{fig-1}(a) and (b) show an optical and SEM image, respectively, of a typical exfoliated hBN flake. The precise flake thickness is measured using a phase-shift interferometer (PSI), which measures the optical path length (OPL) through the flake and converts this using RCWA (rigorous coupled-wave analysis) simulations to the physical thickness. This technique was found to be highly accurate for 2D materials, yet is much faster than the commonly used AFMs \cite{vogl_campbell_buchler_lu_lam_2018,C9NR04269E}. An OPL map is shown in Figure \ref{fig-1}(c), with the physical thickness along the dashed line shown in Figure \ref{fig-1}(d).\\
\indent The quantum emitters are created by irradiating the hBN flake with an electron beam using a scanning electron microscope without any pre-treatment. The high lateral resolution of the SEM allows us to restrict the interaction of the electrons with the hBN to small volumes, thereby localizing the emitter formation at any arbitrarily chosen spots with sub-micron precision. The number of emitters created depends on the fabrication parameters. For all experiments, we used an acceleration voltage of 3 kV and an electron current of 25 pA in the field-free operation mode of the SEM. The theoretical resolution limit of our SEM is 2.8 nm, however, with the used beam alignment we estimate the beam diameter to be around 300 nm. Since this is already the diffraction limit for the subsequent optical characterization, we did not optimize this any further. The analysis of the irradiated spot size is summarized in the Supplementary Section S2. The dose is controlled by the irradiation time and we found an optimal dwell times of 10 s, which results in a particle fluence of $7.7\times 10^{17}$ cm$^{-2}$. Pointing the electron beam to a single point produces single spots as is evident from the SEM image (see Figure \ref{fig-1}(b)), however, higher doses allow us to directly write emitter ensembles and larger patterns (see Supplementary Section S1). We note that the fluence values used to record the SEM images ($10^{13}$ cm$^{-2}$) are low compared to the conditions for emitter writing to avoid accidental fabrication of emitters.\\
\indent It is known from the literature that many fabrication techniques require post-processing, in particular thermal annealing which either activates the emitter or at least enhances the intensity \cite{tran_elbadawi_totonjian_lobo_grosso_moon_englund_ford_aharonovich_toth_et,doi:10.1021/acsphotonics.2c00631, LiXuMendelsonKianiniaTothAharonovich+2019+2049+2055}. To investigate the impact of annealing on the emission properties (which are studied in detail in the next section), a flake is prepared with an emitter array of $10\times 4$ irradiated spots. Our photoluminescence (PL) map reveal that the emitter array that is present initially disappears after treatment with typical annealing conditions in a furnace at 850 $^\circ$C under an inert argon atmosphere (600 mbar pressure) for 30 min (see Supplementary Section S3). We found that this annealing can activate other (uncorrelated) contaminants, but also heals ours through the SEM activated or created emitters. This suggests a high mobility and subsequent instability of the involved defects at such high temperatures. As a consequence, we conclude there is no post-processing necessary after the SEM treatment, which simplifies the fabrication. Other experiments on electron-irradiated hBN emitters have confirmed this \cite{ngoc,naturecomm}. We note that Gale \textit{et al.}\ reported that pre-annealing can enhance cathodoluminescence at 305 nm, which reveals the presence of the C$_{\text{B}}$C$_{\text{N}}$ dimer that can be correlated to the blue emitter \cite{doi:10.1021/acsphotonics.2c00631}.\\
\indent The emitter fabrication with an electron beam is influenced by the electron beam diameter and electron scattering in the hBN lattice. To further study this interaction, an electron dose-dependent irradiation is performed by varying the electron irradiation time. An array of $4 \times 3$ emitters is fabricated where each spot is irradiated for a different irradiation time. The photon count rate at the irradiated spot presents the saturation behaviour as a function of irradiation time. This indicates the creation/activation of the emitter is bounded to the maximum emitter density \cite{Roux2022}. The optical characterization (for the details see the following section) and analysis are summarized in Supplementary Section S4. However, the emission from individual emitters in an ensemble is uncorrelated and therefore, these sites are not emitting single photons anymore. The probability of forming a single isolated emitter decreases with increasing the electron fluence, while at the same time, the probability of multiple emitters increases. This suggests that the emitter's formation could be governed by Poisson statistics. Thus producing single emitters with a high yield requires a careful calibration of electron fluence.

\section{Results}
\subsection{Optical characterization}
\begin{figure*}
    \includegraphics[width = 0.95\textwidth]{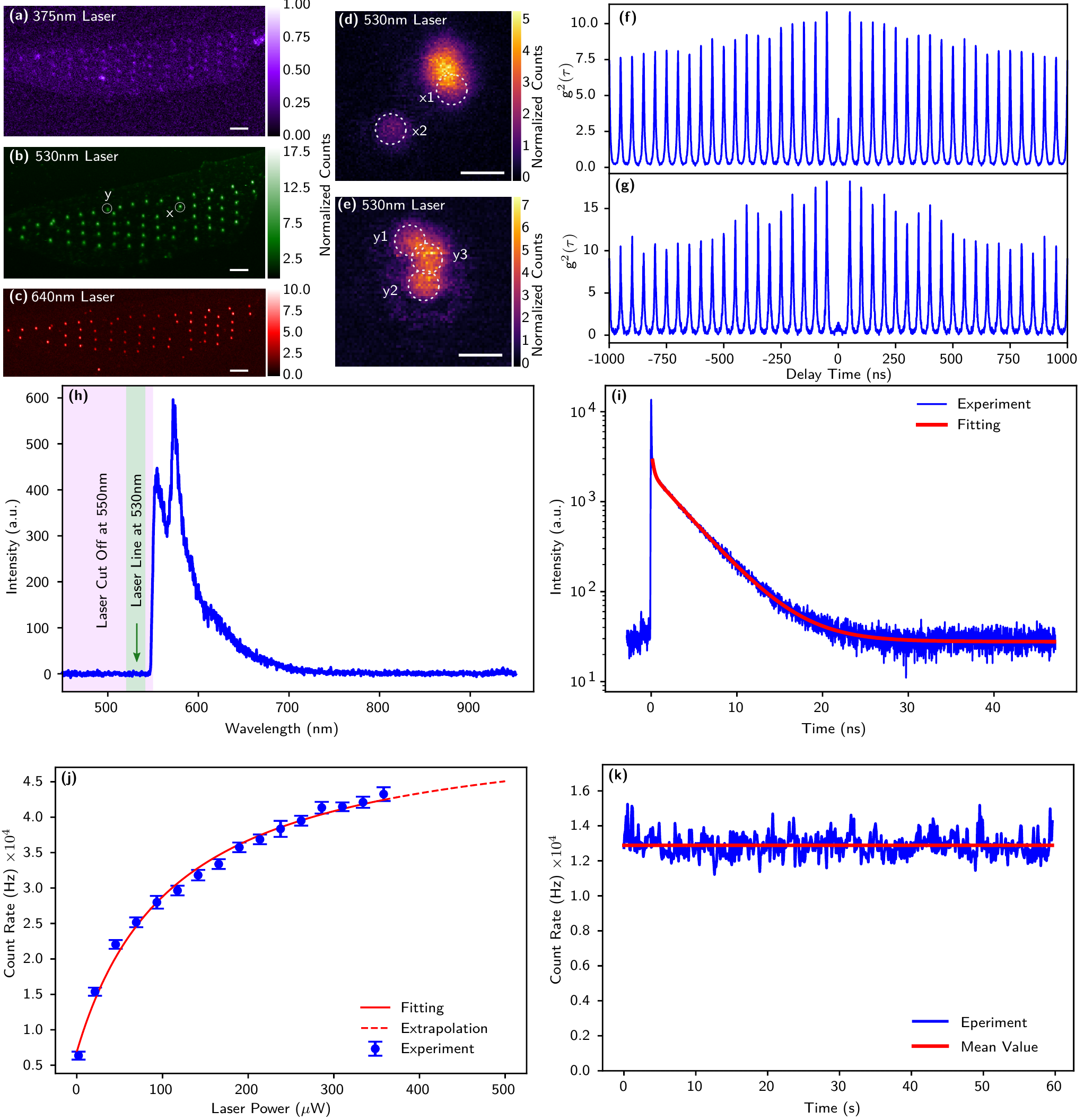}
    \caption{Optical characterization of the emitter array using a time-resolved photoluminescene microscope. (a-c) PL maps of the array excited with a (a) 375 nm laser and a longpass filter at 405 nm, (b) 530 nm laser and a longpass filter at 550 nm, and (c) 640 nm and a bandpass filter transmitting from 650 to 720 nm. Each PL map is normalized to the maximum counts achieved with the 375 nm laser in (a). The white scale bar is 5 $\mu$m. (d,e) PL maps of two individual irradiated spots excited at 530 nm marked with x and y in (b). The spots (x1, x2) and (y1, y2, y3) are individual SPEs with $g^{(2)}(0)$ = (0.536, 0.224) at (x1, x2) and $g^{(2)}(0)$ = (0.379, 0.363, 0.445) at (y1, y2, y3), respectively. The scale bar is 0.5 $\mu$m. The second-order-correlation measurements of y1 and x2 are depicted in (f) and (g), respectively. (h) Spectrum of a sample SPE with a peak of the PL at 575 nm excited at 530 nm and detected with a longpass filter at 550 nm that cuts the emission partially. (i) The time-resolved PL measurement of the SPE reveals an excited-state lifetime of 3.83(1) ns. The initial short peak in our measurement is due to the instrument response function (IRF), which was taken care of by the built-in software while performing the fitting to extract the lifetime of the emitter. (j) The power saturation curve of the emitter is similar to that of a two level system. The saturation intensity of 46.88 kHz and saturation power of 114 $\mu$W is extracted from a fit. (k) The timetrace of a continuously excited emitter over 60 s demonstrates that the emitters exhibit no blinking or bleaching. The mean and standard deviation are 12.9(6) kHz, implying a relative stability of 4.65\,\%.}
    \label{Optical_cha}
\end{figure*}
To identify if the fabrication parameters are feasible, we have optically characterized the created emitters with a commercial fluorescence lifetime imaging microscope (see Supplementary Section S5). All measurements have been performed at room temperature and with an excitation laser pulse with repetition rate of 20 MHz. Our setup allows us to use different laser heads with fixed wavelengths of 375 nm, 530 nm, and 640 nm and pulse lengths in the range below 40 ps to 90 ps. The PL images for the different excitation energies are shown in Figures \ref{Optical_cha}(a-c). As the crystal is always excited below the bandgap, there is bright emission at electron-irradiated spots only. The brightness, however, strongly depends on the excitation energy. Note that the color scale for each PL map has been normalized such that the brightest emitter in Figure \ref{Optical_cha}(a) under excitation with 375 nm laser light has a PL intensity of 1 a.u. As can be seen, the excitation is very inefficient in the UV, in particular, the blue emitters at 435 nm were not present. This suggests that our sample does not contain the C$_{\text{B}}$C$_{\text{N}}$ dimer that can be correlated to the blue emitter \cite{doi:10.1021/acsphotonics.2c00631}. In contrast to the UV excitation, the emitters are the brightest under the illumination of 530 nm excitation laser. We would like to mention that we have dynamically adapted the spacing between the irradiation spots so that our array does not have periodic spacing. This was due to the fact that the mechanically exfoliated flakes typically have random shapes and we did not want to fabricate emitters close to the crystal edges. The mean distance between irradiation spots is roughly 3 $\mu$m. In the overlayed SEM and PL image (see Supplementary Section S6), it becomes clear that all irradiated spots have a corresponding bright spot in the PL image. The number of emitters per irradiated spot varies: while most of the time there is exactly one emitter present, sometimes there is a second emitter nearby (see Figure \ref{Optical_cha}(d)) or emitter ensembles are created in the same spot (see Figure \ref{Optical_cha}(e) and Supplementary Section S7). This is due to the Poisson statistics of the irradiation process. It is also worth mentioning that the bright emission from the irradiated spot is independent of the flake thickness, as shown in the overlayed OPL and PL maps (see Supplementary section S8). This could be due to the fact that the localized electron irradiation depositing carbon atoms which are responsible for the defect formation in the top layer of the hBN.\\
\indent The second-order correlation function $g^{(2)}(\tau)$ of these emitters reveals the presence of single photon emitters with the zero-delay peak under pulsed excitation dropping to 0.36 and 0.22 (without any background correction) in Figure \ref{Optical_cha}(f) and (g), respectively. A $g^{(2)}(0)$ value below 0.5 proves a non-zero overlap with the single photon Fock state \cite{Gr_nwald_2019} and is therefore a commonly-used criterion for single photon emission. We note that only $g^{(2)}(0)=0$ proves rigorously single photon emission, which can be reached when the experimental imperfections are considered \cite{arXiv:2111.01252}. The details about our $g^{(2)}(\tau)$ calculations can be found in Supplementary Section S9. The spectrally-resolved measurement of the emission reveals a peak at 575 nm when the emitter is excited with a 530 nm laser (see Figure \ref{Optical_cha}(h)). 
\\ \indent It is worth noting that our long-pass filter has a cut-off wavelength of 550 nm to suppress the excitation laser at 530 nm which partially suppresses the detection of the emission. We therefore do not have access to the full spectral information about the emitter under 530 nm excitation laser. However, to investigate the possibility of any strong peak emission in the spectrally blocked region below 550 nm, we also used a 470 nm excitation laser to excite the emitters. A long pass filter with a cut-off wavelength at 500 nm is used to collect the spectrum. We do not observe any strong emission when exciting with 470 nm laser, which excludes the possibility of any other emission peak in the blocked spectral region. Moreover, with this excitation laser we also do not observe the 575 nm peak as observed with the 530 nm excitation laser. This could be due to the fact that the 470 nm excitation laser has negligible overlap with the emitter phonon side bands (absorption) of the emitters. This can be estimated by mirroring the spectra around the peak at 575 nm (see Supplementary Section S10). The PL of the emitters under 375 nm excitation was too weak to record a spectrum above the noise floor (confirming inefficient excitation), and under 640 nm excitation, the emitters showed a rather broadband spectrum (probably dominated by phonon coupling).\\
\indent In addition, the peak at 575 nm is near the Raman peak of hBN (the Raman shift of hBN \cite{tran_bray_ford_toth_aharonovich_2016} at 1360 cm$^{-1}$ results in a peak at 571 nm). To determine if the 575 nm peak is the Raman peak we have also measured the temporally-resolved PL as shown in Figure \ref{Optical_cha}(i). A fit (taking into account the instrument response function) reveals an excited state lifetime of 3.83(1) ns, consistent with other reports of hBN quantum emitters \cite{aharonovich_englund_toth_2016, vogl_campbell_buchler_lu_lam_2018, vogl_lecamwasam_buchler_lu_lam_2019}. The spectrally-resolved measurement of the lifetime and $g^{(2)}(\tau)$ function have shown to be independent of the wavelength. We estimated that the 575 nm peak contributes too much to the spectrum to originate from Raman scattering (because this would change the photon statistics and emitter lifetime).\\
\indent The single photon intensity $I$ as a function of the laser excitation power $P$ reveals the typical saturation curve of a two-level system (see Figure \ref{Optical_cha}(j)) and a fit with
\begin{equation}
I(P)=\frac{I_{\mathrm{sat}} P}{P+P_{\mathrm{sat}}}+I_{\mathrm{d}}
\end{equation}
allows us to extract a saturation power of $P_{\mathrm{sat}}=114$~$\mu$W and a maximum photon count rate of $I_{\mathrm{sat}}=46.88$~kHz. The dark count rate $I_{\mathrm{d}}$ has a negligible effect on the fit due to the low dark count rate of our single photon detectors. The detected maximum photon count rate, however, is not the actual maximum emission rate because it is limited by the collection, transmission, and detection efficiencies. The former can be addressed by coupling the emitter with resonant cavities \cite{vogl_lecamwasam_buchler_lu_lam_2019} that shape the emission to be more directional. Interestingly, we have not observed any blinking or bleaching of any of the emitters as shown by our time trace of an emitter in Figure \ref{Optical_cha}(k). While this problem is known for many hBN emitters \cite{Boll:20,doi:10.1021/acsnano.9b00274}, our electron irradiation seems to produce emitters that are not affected by this. \\
\indent We have also investigated the defects for addressable spin states through optically detected magnetic resonance \cite{gottscholl,stern, liu_feng_wang_li_liu_2019}. Our preliminary measurements show no signature of ODMR in the typical bandwidth ranging from 3 GHz to 4 GHz as presented in Supplementary Section S11. However, our present measurements do not exclude the possibility of ODMR signature beyond the mentioned bandwidth, as recent work show the OMDR at 0.7 GHz reported by \citet{stern} and at 2.5 GHz reported by \citet{Mu2022}. Our measurements are carried out on an ensemble of hBN emitters at room temperature. A control measurement with NV centers in diamond, however, demonstrated a significant ODMR contrast of that spin system and thereby, verifying the general functionality of our ODMR setup.

\begin{figure*}
    \includegraphics[width = 1\textwidth]{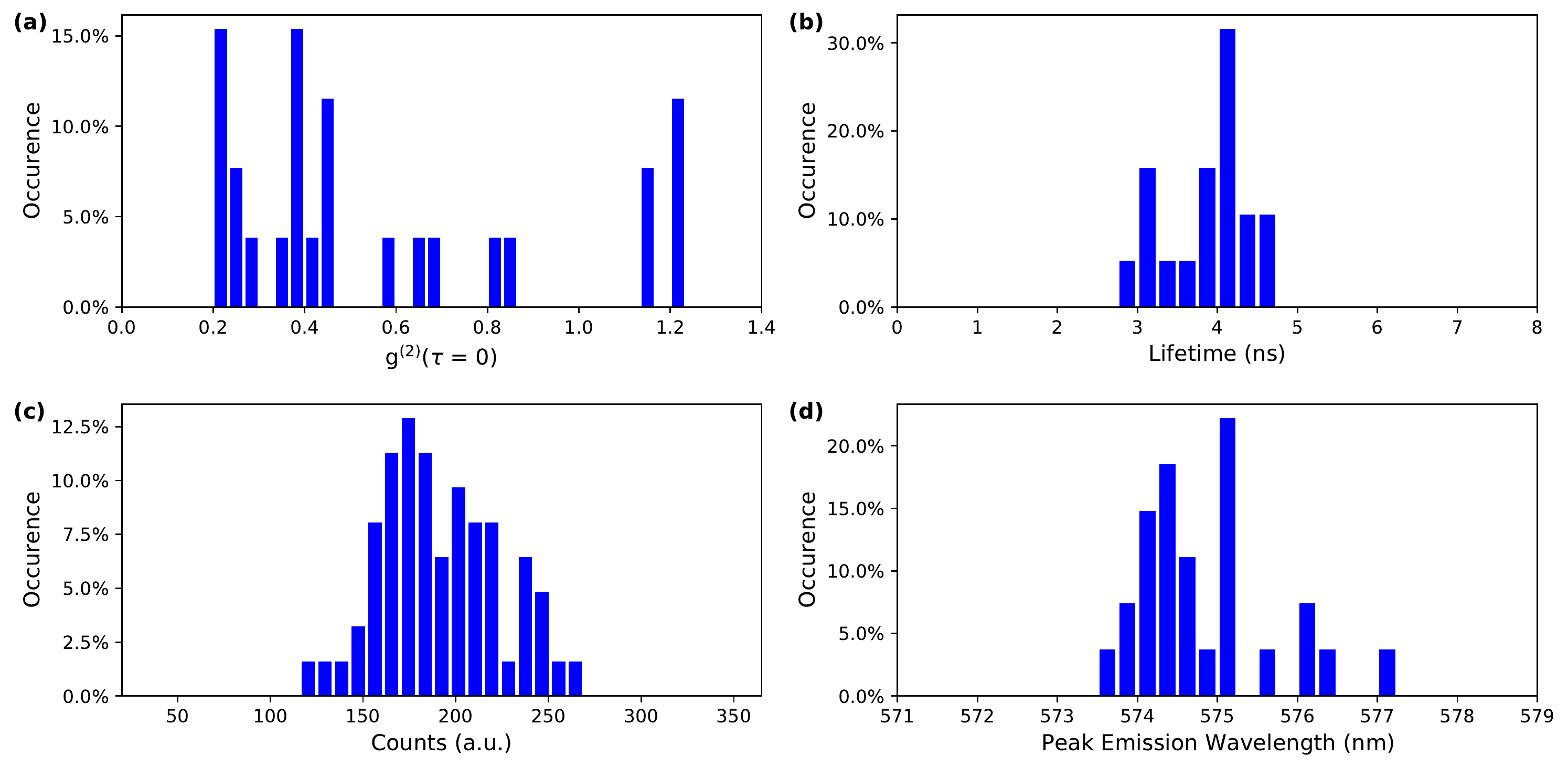}
    \caption{Statistical distribution of photophysical properties from an emitter array on the same flake. (a) The distribution of second-order correlation function at zero time delay $g^{(2)}(0)$ shows a majority of emitters have $g^{(2)}(0)<0.5$ and are therefore classified as single photon emitters. (b) Lifetime distribution of emitters with $g^{(2)}(0)<0.5$ extracted from the correlation function measurements. The mean value and standard deviation is 3.8(5) ns. (c) Brightness distribution of the emitter array excited with the 530 nm laser and a power of 50 $\mu$W. The photon counts are obtained by integration over the spots in the PL map in Figure \ref{Optical_cha}(b), each of which may contain a single emitter or ensembles thereof. (d)  The distribution of peak emission wavelength from the emitters. Gaussian fitting is performed to estimate the peak position. The mean value of the peak emission wavelength is centered around 574.83(84) nm.}
    \label{hist}
\end{figure*}

\begin{figure*}
    \includegraphics[width = 1\textwidth]{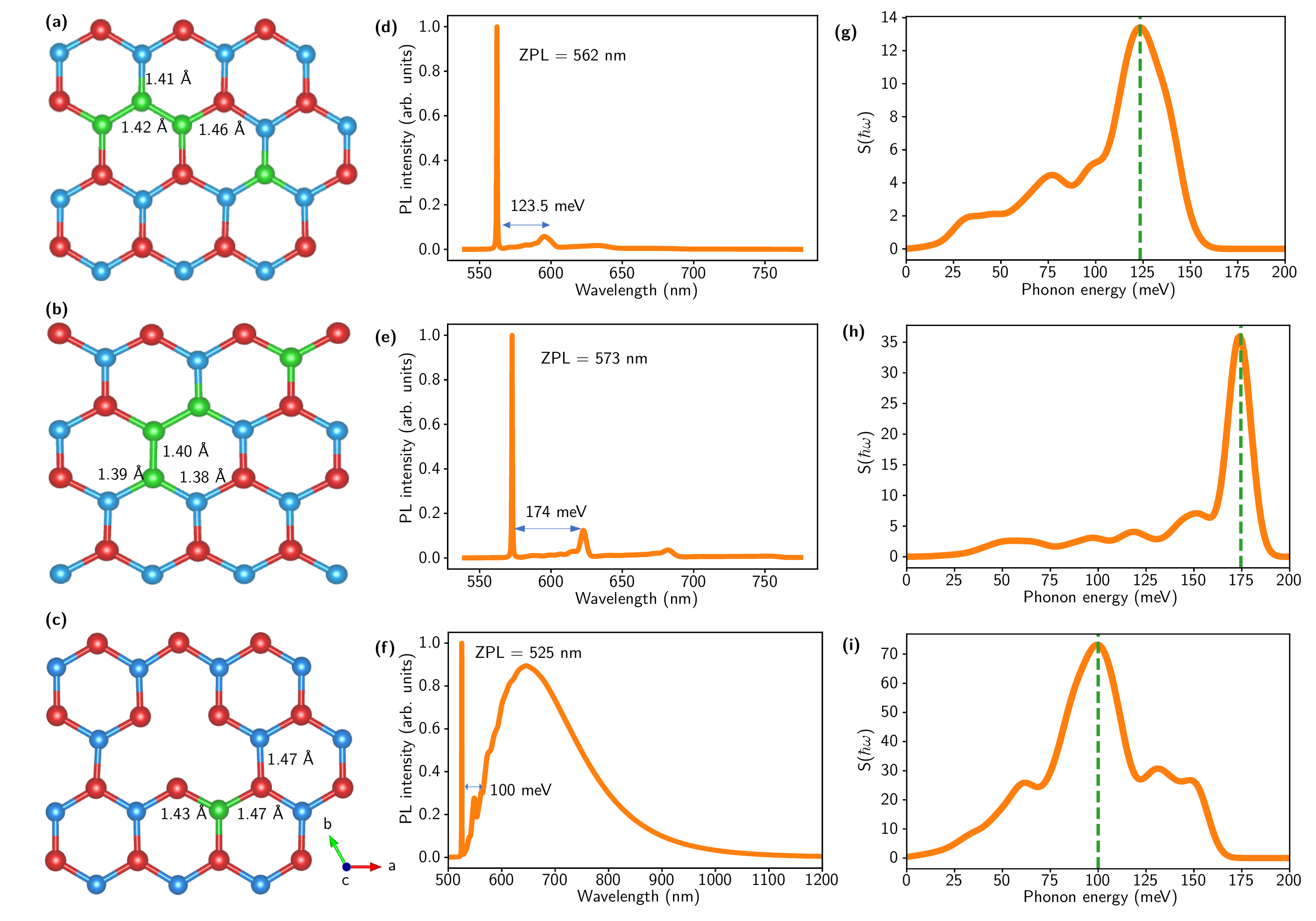}
    \caption{Theoretical simulations. (a-c) Defect structures of potential candidates with the C$_{\text{B}}$C$_{\text{N}}$C$_{\text{B}}$C$_{\text{N}}$ defect in two different configurations in (a) and (b), as well as the C$_\text{N}$V$_\text{N}$ defect in (c). The boron atoms are blue, nitrogen red, and carbon green. The $a$ and $b$ directions are in-plane. (d-f) Theoretical photoluminescence spectra of the defect structures in (a-c), respectively, where the first peak belongs to ZPL while the second peak to the phonon sideband. The Huang-Rhys factor in (g-i) shows the corresponding contributions of the specific phonon modes to the PL spectra. The dashed line marks the peak of the phonon sideband. The broad PL emission in the near infrared in (f) is explained in the main text.}
    \label{fig:PL_DFT}
\end{figure*}

\subsection{Statistical analysis}
To understand if our process also produces high-quality emitters in general, we have analyzed the statistics over a larger data set containing 63 emitters fabricated on the same flake. We have decided to only include emitters hosted on the same flake in this analysis to remove the influence of the host crystal (e.g.\ through local strain) on the emitter properties. Figure \ref{hist}(a) shows the distribution of the $g^{(2)}(0)$ values, which reveals that most of the emitters can be classified as singles, while a few have $g^{(2)}(0)>1$ and are likely ensembles. The PL maps of a subset of these emitters are presented in Supplementary Section S7. 
\\ \indent The histogram of the lifetimes (see Figure \ref{hist}(b)) reveals a mean lifetime of 3.8(5) ns (only counting emitters with $g^{(2)}(0)\leq0.5$). The consistency of the lifetime provides further evidence that the emitters produced are always the same without any additional decay channels for the excited state. We note that we extracted the lifetime for the histogram from the pulsed measurement of the second-order correlation function (see Supplementary Section S12). In the limit of zero excitation power, this method would yield the correct lifetime. In our case we used this method to reduce the measurement time, as we can measure lifetime and second-order correlation time simultaneously. 
\\ \indent The statistical distribution of the emitter brightness is presented in Figure \ref{hist}(c), extracted from the PL map. For all 63 emitters presented in Figure \ref{Optical_cha}(a), we wrote an automated algorithm that detects the emitter position and integrates over a diffraction-limited spot around its center (marked by the red boxes in Supplementary Section S13). 
\\ \indent We also measure the spectrum from various other emitters exhibiting a dip in the second-order correlation measurements. We performed Gaussian fitting to extract the peak emission wavelength, which is bunching around 574.8 nm as shown in the Figure.\ \ref{hist}(d). The mean FWHM of the peak emission wavelength is around 19.56 nm with the standard deviation of 4.44 nm (See Supplementary Section S14). This also confirms the spectral re-producibility of our fabrication method.

\subsection*{Emitter modeling with density functional theory}
To trace back possible defect candidates that could be responsible for the yellow emission, we carried out theoretical studies using density functional theory with the HSE06 functional implemented in Vienna Ab initio Simulation Package (VASP) \cite{vasp1,vasp2} (see details in Supplementary Section S15). It is known that carbon contaminants are bonded to the surface during SEM imaging, which causes the typical dark frames. This suggests that carbon might be involved in the defect, as other calculations and experiments have demonstrated already \cite{Mackoit2019,Auburger2021,Mendelson2021,Cholsuk2022,Jara2020,Li2022}. Since the experimental spectrum as shown in Figure \ref{Optical_cha}(h) is a characteristic signature of all defects, we calculated the PL lineshape by employing the $\Delta$SCF method to determine the spectra of neutral and singly charged carbon complexes listed in Supplementary Section S15.\\
\indent Our simulations identify the C$_{\text{B}}$C$_{\text{N}}$C$_{\text{B}}$C$_{\text{N}}$ and C$_\text{N}$V$_\text{N}$ complexes with the structural configurations depicted in Figure \ref{fig:PL_DFT}(a-c) as potential candidates for carbon-related emitters with a transition energy (i.e.\ ZPL) in the yellow region. In addition to the ZPL, we also calculated the phonon sideband (PSB) to obtain the complete PL spectrum (see Figure \ref{fig:PL_DFT}(d-f)). The peaks can be identified by considering the spectral function of the electron-phonon coupling or Huang-Rhys (HR) factor as shown in Figure \ref{fig:PL_DFT}(g-i). The PSB peaks in the range of 100 meV to 175 meV, which is the typical range for hBN quantum emitters. Notably, the PL spectra of the C$_\text{N}$V$_\text{N}$ defect exhibits a broad peak at 380 meV from the ZPL in the calculated PL spectrum. This phenomenon can be justified by the fact that these modes have a higher energy ($>200$ meV) than the zone-center longitudinal optical phonon mode~(LO($\Gamma$)) \cite{Vuong2016}. As coupling to electrons scales with the inverse of the phonon wave vector from the Fr\"ohlich interaction, the electron-phonon matrix element will subsequently diverge due to the singularity at this point and lead to broad PL. Based on the HR factor, the assignment of the ZPL becomes evident. To identify the contributions to the ZPL, the HR factor yielded small values of 1.033, 1.372 and 4.795, corresponding to the configuration coordinates between ground and excited states shifted by 0.456 amu$^{1/2}$\AA, 0.435 amu$^{1/2}$\AA, and 0.994 amu$^{1/2}$\AA, respectively, for the defect complexes shown in Figure \ref{fig:PL_DFT}(a-c). This reveals that the defects inherently emit at their assigned ZPL with only a weak contribution from phonon coupling in case of the C$_{\text{B}}$C$_{\text{N}}$C$_{\text{B}}$C$_{\text{N}}$ configurations and with strong phonon coupling for the C$_{\text{N}}$V$_{\text{N}}$ defect. This DFT calculation consequently proposes that C$_{\text{B}}$C$_{\text{N}}$C$_{\text{B}}$C$_{\text{N}}$ is one of the potential candidates responsible for this range of emission, whereas C$_{\text{N}}$V$_{\text{N}}$ is unlikely due to the mismatch of the PSB with the experiment.\\
\indent Moreover, when comparing among singlet, doublet, and triplet spin configurations of C$_{\text{B}}$C$_{\text{N}}$C$_{\text{B}}$C$_{\text{N}}$, it favors the singlet configuration, implying the absence of a meta-stable triplet state. C$_{\text{B}}$C$_{\text{N}}$C$_{\text{B}}$C$_{\text{N}}$ is therefore unlikely to exhibit any ODMR signature, consistent with the experimental result. Our DFT calculations achieved to narrow down the possible defect candidates with ZPL relevant to the yellow region, however, we note that the broad emission observed in the experiment still requires further analysis, which is beyond the scope of this work. \\
\indent We finally answer the question of the formation mechanism. Electrons with an energy of 3 keV do not have enough momentum for knock-on damage and to displace atoms. It is likely that the electrons can bond carbon to the hBN (which causes the black spots in the SEM image) and then break these bonds and re-organize them to form the carbon complexes. Another possibility is the change of the charge state of already present defects or the enabling of a single photon emission mechanism based on donor-acceptor pairs \cite{doi:10.1021/acs.nanolett.1c04647}.

\section{Conclusions}
We have demonstrated the localized fabrication of SPEs in exfoliated multilayers of hBN using focused electron beam irradiation with a well-defined set of emission properties and a high yield. Our process does not require any pre- or post-processing (e.g.\ thermal annealing) of the hBN crystals. All fabricated color centers emit consistently at 575 nm (2.16 eV). This has not been observed in previous work fabricating emitters with reproducible blue emission using similar electron irradiation \cite{ngoc,naturecomm}. We believe that this could be due to our sample not containing the carbon dimer C$_{\text{B}}$C$_{\text{N}}$ that was linked to this emission \cite{doi:10.1021/acsphotonics.2c00631}. We speculate that the origin of our emitter is due to the deposition of carbon by the electron beam irradiation, the breaking and re-organization of chemical bonds, the change of the charge state of existing defects (i.e., activating them optically), or the enabling of single photon emission based on donor-acceptor pairs.\cite{doi:10.1021/acs.nanolett.1c04647}. This does not, however, provide an explanation why previous other works have not observed our yellow emitters. For a better overview of this work, we also provide a table comparing the recent development of 2D materials based single photon emitters in Supplementary Section S16.\\
\indent Our ODMR measurements have not revealed any addressable spin states. The theoretical analysis using DFT calculations allows us to propose possible atomic structures, namely of the carbon complex C$_{\text{B}}$C$_{\text{N}}$C$_{\text{B}}$C$_{\text{N}}$. The process is reproducible and repeatable and thus allows us to fabricate identical SPEs on demand and with deep sub-micron placement accuracy. Furthermore, our study also opens up a new way to understand the nature and photophysical properties of the defect-based emitters in hBN. As the method is scalable and compatible with coupling to nanophotonics, we expect use in near-future quantum technology applications. It was recently pointed out that hBN emitters have naturally spectral overlap with many important wavelengths for quantum technologies \cite{Cholsuk2022}. When enhanced with resonators that reduce the linewidth \cite{vogl_lecamwasam_buchler_lu_lam_2019}, the yellow emitters presented here could be used to couple to other solid-state quantum systems in quantum networks such as to the PbV$^-$ color center in diamond with its ZPL at 552 nm or to the sodium D transitions at 589 nm in alkali vapor-based quantum memories \cite{Cholsuk2022}. Such usage, however, is only possible if the quantum emitters can be fabricated in a here demonstrated reproducible fashion.

\section*{Supplementary Material}
Supplementary Material is available for this manuscript.

\section*{Conflict of Interest}
The authors have no conflicts to disclose.

\section*{Author contributions}
A.K. and M.M. prepared the samples. A.K. performed the irradiation. A.K. and T.M. carried out the optical characterization. C.C. and S.S. performed the DFT calculations. A.Z. carried out the ODMR experiments. A.K. and C.C. analyzed the data. F.E., S.S., and T.V. supervised the work. T.V. conceived the project.

\begin{acknowledgments}
This work was funded by the Deutsche Forschungsgemeinschaft (DFG, German Research Foundation) - Projektnummer 445275953. The authors acknowledge support by the German Space Agency DLR with funds provided by the Federal Ministry for Economic Affairs and and Climate Action BMWK under grant number 50WM2165 (QUICK3). T.V. and F.E. are funded by the Federal Ministry of Education and Research (BMBF) under grant number 13N16292 and 13XP5053A, respectively. The major instrumentation used in this work was funded by the Free State of Thuringia via the projects 2015 FOR 0005 (ACP-FIB) and 2017 IZN 0012 (InQuoSens). C.C. acknowledges a Development and Promotion of Science and Technology Talents Project (DPST) scholarship by the Royal Thai Government. S.S. acknowledges financial support from the NSRF via the Program Management Unit for Human Resources \& Institutional Development, Research and Innovation (grant number B05F650024). The computational experiments were supported by resources of the Friedrich Schiller University Jena supported in part by DFG grants INST 275/334-1 FUGG and INST 275/363-1 FUGG. We thank Jeetendra Gour for the fabrication of the cross grid patterned substrate and acknowledge fruitful discussions with Igor Aharonovich on the potential assignment of the Raman peak.
\end{acknowledgments}

\section*{Data sharing policy}
The data that support the findings of this study are available from the corresponding author upon reasonable request.

\bibliography{main}

%aipnum4-2.bst 2019-01-14 (MD) hand-edited version of apsrev4-1.bst
\providecommand{\noopsort}[1]{}\providecommand{\singleletter}[1]{#1}%
\begin{thebibliography}{60}%
\makeatletter
\providecommand \@ifxundefined [1]{%
 \@ifx{#1\undefined}
}%
\providecommand \@ifnum [1]{%
 \ifnum #1\expandafter \@firstoftwo
 \else \expandafter \@secondoftwo
 \fi
}%
\providecommand \@ifx [1]{%
 \ifx #1\expandafter \@firstoftwo
 \else \expandafter \@secondoftwo
 \fi
}%
\providecommand \natexlab [1]{#1}%
\providecommand \enquote  [1]{``#1''}%
\providecommand \bibnamefont  [1]{#1}%
\providecommand \bibfnamefont [1]{#1}%
\providecommand \citenamefont [1]{#1}%
\providecommand \href@noop [0]{\@secondoftwo}%
\providecommand \href [0]{\begingroup \@sanitize@url \@href}%
\providecommand \@href[1]{\@@startlink{#1}\@@href}%
\providecommand \@@href[1]{\endgroup#1\@@endlink}%
\providecommand \@sanitize@url [0]{\catcode `\\12\catcode `\$12\catcode
  `\&12\catcode `\#12\catcode `\^12\catcode `\_12\catcode `\%12\relax}%
\providecommand \@@startlink[1]{}%
\providecommand \@@endlink[0]{}%
\providecommand \url  [0]{\begingroup\@sanitize@url \@url }%
\providecommand \@url [1]{\endgroup\@href {#1}{\urlprefix }}%
\providecommand \urlprefix  [0]{URL }%
\providecommand \Eprint [0]{\href }%
\providecommand \doibase [0]{https://doi.org/}%
\providecommand \selectlanguage [0]{\@gobble}%
\providecommand \bibinfo  [0]{\@secondoftwo}%
\providecommand \bibfield  [0]{\@secondoftwo}%
\providecommand \translation [1]{[#1]}%
\providecommand \BibitemOpen [0]{}%
\providecommand \bibitemStop [0]{}%
\providecommand \bibitemNoStop [0]{.\EOS\space}%
\providecommand \EOS [0]{\spacefactor3000\relax}%
\providecommand \BibitemShut  [1]{\csname bibitem#1\endcsname}%
\let\auto@bib@innerbib\@empty
%</preamble>
\bibitem [{\citenamefont {O'Brien}, \citenamefont {Furusawa},\ and\
  \citenamefont {Vučković}(2009)}]{obrien_furusawa_vuckovic_2009}%
  \BibitemOpen
  \bibfield  {author} {\bibinfo {author} {\bibfnamefont {J.~L.}\ \bibnamefont
  {O'Brien}}, \bibinfo {author} {\bibfnamefont {A.}~\bibnamefont {Furusawa}},\
  and\ \bibinfo {author} {\bibfnamefont {J.}~\bibnamefont {Vučković}},\
  }\bibfield  {title} {\enquote {\bibinfo {title} {Photonic quantum
  technologies},}\ }\href {https://doi.org/10.1038/nphoton.2009.229} {\bibfield
   {journal} {\bibinfo  {journal} {Nat. Photon.}\ }\textbf {\bibinfo {volume}
  {3}},\ \bibinfo {pages} {687–695} (\bibinfo {year} {2009})}\BibitemShut
  {NoStop}%
\bibitem [{\citenamefont {Mukai}\ and\ \citenamefont
  {Hatano}(2020)}]{PhysRevResearch.2.023378}%
  \BibitemOpen
  \bibfield  {author} {\bibinfo {author} {\bibfnamefont {K.}~\bibnamefont
  {Mukai}}\ and\ \bibinfo {author} {\bibfnamefont {N.}~\bibnamefont {Hatano}},\
  }\bibfield  {title} {\enquote {\bibinfo {title} {Discrete-time quantum walk
  on complex networks for community detection},}\ }\href
  {https://doi.org/10.1103/PhysRevResearch.2.023378} {\bibfield  {journal}
  {\bibinfo  {journal} {Phys. Rev. Res.}\ }\textbf {\bibinfo {volume} {2}},\
  \bibinfo {pages} {023378} (\bibinfo {year} {2020})}\BibitemShut {NoStop}%
\bibitem [{\citenamefont {Aspuru-Guzik}\ and\ \citenamefont
  {Walther}(2012)}]{aspuru-guzik_walther_2012}%
  \BibitemOpen
  \bibfield  {author} {\bibinfo {author} {\bibfnamefont {A.}~\bibnamefont
  {Aspuru-Guzik}}\ and\ \bibinfo {author} {\bibfnamefont {P.}~\bibnamefont
  {Walther}},\ }\bibfield  {title} {\enquote {\bibinfo {title} {Photonic
  quantum simulators},}\ }\href {https://doi.org/10.1038/nphys2253} {\bibfield
  {journal} {\bibinfo  {journal} {Nat. Phys.}\ }\textbf {\bibinfo {volume}
  {8}},\ \bibinfo {pages} {285–291} (\bibinfo {year} {2012})}\BibitemShut
  {NoStop}%
\bibitem [{\citenamefont {Cai}\ \emph {et~al.}(2013)\citenamefont {Cai},
  \citenamefont {Retzker}, \citenamefont {Jelezko},\ and\ \citenamefont
  {Plenio}}]{cai_retzker_jelezko_plenio_2013}%
  \BibitemOpen
  \bibfield  {author} {\bibinfo {author} {\bibfnamefont {J.}~\bibnamefont
  {Cai}}, \bibinfo {author} {\bibfnamefont {A.}~\bibnamefont {Retzker}},
  \bibinfo {author} {\bibfnamefont {F.}~\bibnamefont {Jelezko}},\ and\ \bibinfo
  {author} {\bibfnamefont {M.~B.}\ \bibnamefont {Plenio}},\ }\bibfield  {title}
  {\enquote {\bibinfo {title} {A large-scale quantum simulator on a diamond
  surface at room temperature},}\ }\href {https://doi.org/10.1038/nphys2519}
  {\bibfield  {journal} {\bibinfo  {journal} {Nat. Phys.}\ }\textbf {\bibinfo
  {volume} {9}},\ \bibinfo {pages} {168–173} (\bibinfo {year}
  {2013})}\BibitemShut {NoStop}%
\bibitem [{\citenamefont {Lvovsky}, \citenamefont {Sanders},\ and\
  \citenamefont {Tittel}(2009)}]{lvovsky_sanders_tittel_2009}%
  \BibitemOpen
  \bibfield  {author} {\bibinfo {author} {\bibfnamefont {A.~I.}\ \bibnamefont
  {Lvovsky}}, \bibinfo {author} {\bibfnamefont {B.~C.}\ \bibnamefont
  {Sanders}},\ and\ \bibinfo {author} {\bibfnamefont {W.}~\bibnamefont
  {Tittel}},\ }\bibfield  {title} {\enquote {\bibinfo {title} {Optical quantum
  memory},}\ }\href {https://doi.org/10.1038/nphoton.2009.231} {\bibfield
  {journal} {\bibinfo  {journal} {Nat. Photon.}\ }\textbf {\bibinfo {volume}
  {3}},\ \bibinfo {pages} {706–714} (\bibinfo {year} {2009})}\BibitemShut
  {NoStop}%
\bibitem [{\citenamefont {Lo}, \citenamefont {Curty},\ and\ \citenamefont
  {Tamaki}(2014)}]{Lo_2014}%
  \BibitemOpen
  \bibfield  {author} {\bibinfo {author} {\bibfnamefont {H.-K.}\ \bibnamefont
  {Lo}}, \bibinfo {author} {\bibfnamefont {M.}~\bibnamefont {Curty}},\ and\
  \bibinfo {author} {\bibfnamefont {K.}~\bibnamefont {Tamaki}},\ }\bibfield
  {title} {\enquote {\bibinfo {title} {Secure quantum key distribution},}\
  }\href {https://doi.org/10.1038/nphoton.2014.149} {\bibfield  {journal}
  {\bibinfo  {journal} {Nat. Photon.}\ }\textbf {\bibinfo {volume} {8}},\
  \bibinfo {pages} {595--604} (\bibinfo {year} {2014})}\BibitemShut {NoStop}%
\bibitem [{\citenamefont {Senellart}, \citenamefont {Solomon},\ and\
  \citenamefont {White}(2017)}]{10.1038/nnano.2017.218}%
  \BibitemOpen
  \bibfield  {author} {\bibinfo {author} {\bibfnamefont {P.}~\bibnamefont
  {Senellart}}, \bibinfo {author} {\bibfnamefont {G.}~\bibnamefont {Solomon}},\
  and\ \bibinfo {author} {\bibfnamefont {A.}~\bibnamefont {White}},\ }\bibfield
   {title} {\enquote {\bibinfo {title} {High-performance semiconductor
  quantum-dot single-photon sources},}\ }\href
  {https://doi.org/10.1038/nnano.2017.218} {\bibfield  {journal} {\bibinfo
  {journal} {Nat. Nanotechnol.}\ }\textbf {\bibinfo {volume} {12}},\ \bibinfo
  {pages} {1026--1039} (\bibinfo {year} {2017})}\BibitemShut {NoStop}%
\bibitem [{\citenamefont {Aharonovich}, \citenamefont {Englund},\ and\
  \citenamefont {Toth}(2016)}]{aharonovich_englund_toth_2016}%
  \BibitemOpen
  \bibfield  {author} {\bibinfo {author} {\bibfnamefont {I.}~\bibnamefont
  {Aharonovich}}, \bibinfo {author} {\bibfnamefont {D.}~\bibnamefont
  {Englund}},\ and\ \bibinfo {author} {\bibfnamefont {M.}~\bibnamefont
  {Toth}},\ }\bibfield  {title} {\enquote {\bibinfo {title} {Solid-state
  single-photon emitters},}\ }\href {https://doi.org/10.1038/nphoton.2016.186}
  {\bibfield  {journal} {\bibinfo  {journal} {Nat. Photon.}\ }\textbf {\bibinfo
  {volume} {10}},\ \bibinfo {pages} {631–641} (\bibinfo {year}
  {2016})}\BibitemShut {NoStop}%
\bibitem [{\citenamefont {Vogl}\ \emph
  {et~al.}(2019{\natexlab{a}})\citenamefont {Vogl}, \citenamefont {Doherty},
  \citenamefont {Buchler}, \citenamefont {Lu},\ and\ \citenamefont
  {Lam}}]{C9NR04269E}%
  \BibitemOpen
  \bibfield  {author} {\bibinfo {author} {\bibfnamefont {T.}~\bibnamefont
  {Vogl}}, \bibinfo {author} {\bibfnamefont {M.~W.}\ \bibnamefont {Doherty}},
  \bibinfo {author} {\bibfnamefont {B.~C.}\ \bibnamefont {Buchler}}, \bibinfo
  {author} {\bibfnamefont {Y.}~\bibnamefont {Lu}},\ and\ \bibinfo {author}
  {\bibfnamefont {P.~K.}\ \bibnamefont {Lam}},\ }\bibfield  {title} {\enquote
  {\bibinfo {title} {Atomic localization of quantum emitters in multilayer
  hexagonal boron nitride},}\ }\href {https://doi.org/10.1039/C9NR04269E}
  {\bibfield  {journal} {\bibinfo  {journal} {Nanoscale}\ }\textbf {\bibinfo
  {volume} {11}},\ \bibinfo {pages} {14362--14371} (\bibinfo {year}
  {2019}{\natexlab{a}})}\BibitemShut {NoStop}%
\bibitem [{\citenamefont {Hoese}\ \emph {et~al.}(2020)\citenamefont {Hoese},
  \citenamefont {Reddy}, \citenamefont {Dietrich}, \citenamefont {Koch},
  \citenamefont {Fehler}, \citenamefont {Doherty},\ and\ \citenamefont
  {Kubanek}}]{doi:10.1126/sciadv.aba6038}%
  \BibitemOpen
  \bibfield  {author} {\bibinfo {author} {\bibfnamefont {M.}~\bibnamefont
  {Hoese}}, \bibinfo {author} {\bibfnamefont {P.}~\bibnamefont {Reddy}},
  \bibinfo {author} {\bibfnamefont {A.}~\bibnamefont {Dietrich}}, \bibinfo
  {author} {\bibfnamefont {M.~K.}\ \bibnamefont {Koch}}, \bibinfo {author}
  {\bibfnamefont {K.~G.}\ \bibnamefont {Fehler}}, \bibinfo {author}
  {\bibfnamefont {M.~W.}\ \bibnamefont {Doherty}},\ and\ \bibinfo {author}
  {\bibfnamefont {A.}~\bibnamefont {Kubanek}},\ }\bibfield  {title} {\enquote
  {\bibinfo {title} {Mechanical decoupling of quantum emitters in hexagonal
  boron nitride from low-energy phonon modes},}\ }\href
  {https://doi.org/10.1126/sciadv.aba6038} {\bibfield  {journal} {\bibinfo
  {journal} {Sci. Adv.}\ }\textbf {\bibinfo {volume} {6}},\ \bibinfo {pages}
  {eaba6038} (\bibinfo {year} {2020})}\BibitemShut {NoStop}%
\bibitem [{\citenamefont {Hoese}\ \emph {et~al.}(2022)\citenamefont {Hoese},
  \citenamefont {Koch}, \citenamefont {Breuning}, \citenamefont {Lettner},
  \citenamefont {Fehler},\ and\ \citenamefont
  {Kubanek}}]{doi:10.1063/5.0074946}%
  \BibitemOpen
  \bibfield  {author} {\bibinfo {author} {\bibfnamefont {M.}~\bibnamefont
  {Hoese}}, \bibinfo {author} {\bibfnamefont {M.~K.}\ \bibnamefont {Koch}},
  \bibinfo {author} {\bibfnamefont {F.}~\bibnamefont {Breuning}}, \bibinfo
  {author} {\bibfnamefont {N.}~\bibnamefont {Lettner}}, \bibinfo {author}
  {\bibfnamefont {K.~G.}\ \bibnamefont {Fehler}},\ and\ \bibinfo {author}
  {\bibfnamefont {A.}~\bibnamefont {Kubanek}},\ }\bibfield  {title} {\enquote
  {\bibinfo {title} {Single photon randomness originating from the symmetric
  dipole emission pattern of quantum emitters},}\ }\href
  {https://doi.org/10.1063/5.0074946} {\bibfield  {journal} {\bibinfo
  {journal} {Appl. Phys. Lett.}\ }\textbf {\bibinfo {volume} {120}},\ \bibinfo
  {pages} {044001} (\bibinfo {year} {2022})}\BibitemShut {NoStop}%
\bibitem [{\citenamefont {Klein}\ \emph {et~al.}(2019)\citenamefont {Klein},
  \citenamefont {Lorke}, \citenamefont {Florian}, \citenamefont {Sigger},
  \citenamefont {Sigl}, \citenamefont {Rey}, \citenamefont {Wierzbowski},
  \citenamefont {Cerne}, \citenamefont {Müller}, \citenamefont
  {Mitterreiter},\ and\ \citenamefont
  {et~al.}}]{klein_lorke_florian_sigger_sigl_rey_wierzbowski_cerne_muller_mitterreiter_et}%
  \BibitemOpen
  \bibfield  {author} {\bibinfo {author} {\bibfnamefont {J.}~\bibnamefont
  {Klein}}, \bibinfo {author} {\bibfnamefont {M.}~\bibnamefont {Lorke}},
  \bibinfo {author} {\bibfnamefont {M.}~\bibnamefont {Florian}}, \bibinfo
  {author} {\bibfnamefont {F.}~\bibnamefont {Sigger}}, \bibinfo {author}
  {\bibfnamefont {L.}~\bibnamefont {Sigl}}, \bibinfo {author} {\bibfnamefont
  {S.}~\bibnamefont {Rey}}, \bibinfo {author} {\bibfnamefont {J.}~\bibnamefont
  {Wierzbowski}}, \bibinfo {author} {\bibfnamefont {J.}~\bibnamefont {Cerne}},
  \bibinfo {author} {\bibfnamefont {K.}~\bibnamefont {Müller}}, \bibinfo
  {author} {\bibfnamefont {E.}~\bibnamefont {Mitterreiter}},\ and\ \bibinfo
  {author} {\bibnamefont {et~al.}},\ }\bibfield  {title} {\enquote {\bibinfo
  {title} {Site-selectively generated photon emitters in monolayer {MoS2} via
  local helium ion irradiation},}\ }\href
  {https://doi.org/10.1038/s41467-019-10632-z} {\bibfield  {journal} {\bibinfo
  {journal} {Nat. Commun.}\ }\textbf {\bibinfo {volume} {10}},\ \bibinfo
  {pages} {2755} (\bibinfo {year} {2019})}\BibitemShut {NoStop}%
\bibitem [{\citenamefont {Klein}\ \emph {et~al.}(2021)\citenamefont {Klein},
  \citenamefont {Sigl}, \citenamefont {Gyger}, \citenamefont {Barthelmi},
  \citenamefont {Florian}, \citenamefont {Rey}, \citenamefont {Taniguchi},
  \citenamefont {Watanabe}, \citenamefont {Jahnke}, \citenamefont {Kastl},\
  and\ \citenamefont
  {et~al.}}]{klein_sigl_gyger_barthelmi_florian_rey_taniguchi_watanabe_jahnke_kastl_et}%
  \BibitemOpen
  \bibfield  {author} {\bibinfo {author} {\bibfnamefont {J.}~\bibnamefont
  {Klein}}, \bibinfo {author} {\bibfnamefont {L.}~\bibnamefont {Sigl}},
  \bibinfo {author} {\bibfnamefont {S.}~\bibnamefont {Gyger}}, \bibinfo
  {author} {\bibfnamefont {K.}~\bibnamefont {Barthelmi}}, \bibinfo {author}
  {\bibfnamefont {M.}~\bibnamefont {Florian}}, \bibinfo {author} {\bibfnamefont
  {S.}~\bibnamefont {Rey}}, \bibinfo {author} {\bibfnamefont {T.}~\bibnamefont
  {Taniguchi}}, \bibinfo {author} {\bibfnamefont {K.}~\bibnamefont {Watanabe}},
  \bibinfo {author} {\bibfnamefont {F.}~\bibnamefont {Jahnke}}, \bibinfo
  {author} {\bibfnamefont {C.}~\bibnamefont {Kastl}},\ and\ \bibinfo {author}
  {\bibnamefont {et~al.}},\ }\bibfield  {title} {\enquote {\bibinfo {title}
  {Engineering the luminescence and generation of individual defect emitters in
  atomically thin mos2},}\ }\href
  {https://doi.org/10.1021/acsphotonics.0c01907} {\bibfield  {journal}
  {\bibinfo  {journal} {ACS Photonics}\ }\textbf {\bibinfo {volume} {8}},\
  \bibinfo {pages} {669–677} (\bibinfo {year} {2021})}\BibitemShut {NoStop}%
\bibitem [{\citenamefont {Srivastava}\ \emph {et~al.}(2015)\citenamefont
  {Srivastava}, \citenamefont {Sidler}, \citenamefont {Allain}, \citenamefont
  {Lembke}, \citenamefont {Kis},\ and\ \citenamefont
  {Imamoğlu}}]{srivastava_sidler_allain_lembke_kis_imamoglu_2015}%
  \BibitemOpen
  \bibfield  {author} {\bibinfo {author} {\bibfnamefont {A.}~\bibnamefont
  {Srivastava}}, \bibinfo {author} {\bibfnamefont {M.}~\bibnamefont {Sidler}},
  \bibinfo {author} {\bibfnamefont {A.~V.}\ \bibnamefont {Allain}}, \bibinfo
  {author} {\bibfnamefont {D.~S.}\ \bibnamefont {Lembke}}, \bibinfo {author}
  {\bibfnamefont {A.}~\bibnamefont {Kis}},\ and\ \bibinfo {author}
  {\bibfnamefont {A.}~\bibnamefont {Imamoğlu}},\ }\bibfield  {title} {\enquote
  {\bibinfo {title} {Optically active quantum dots in monolayer wse2},}\ }\href
  {https://doi.org/10.1038/nnano.2015.60} {\bibfield  {journal} {\bibinfo
  {journal} {Nat. Nanotechnol.}\ }\textbf {\bibinfo {volume} {10}},\ \bibinfo
  {pages} {491–496} (\bibinfo {year} {2015})}\BibitemShut {NoStop}%
\bibitem [{\citenamefont {He}\ \emph {et~al.}(2015)\citenamefont {He},
  \citenamefont {Clark}, \citenamefont {Schaibley}, \citenamefont {He},
  \citenamefont {Chen}, \citenamefont {Wei}, \citenamefont {Ding},
  \citenamefont {Zhang}, \citenamefont {Yao}, \citenamefont {Xu},\ and\
  \citenamefont
  {et~al.}}]{he_clark_schaibley_he_chen_wei_ding_zhang_yao_xu_et}%
  \BibitemOpen
  \bibfield  {author} {\bibinfo {author} {\bibfnamefont {Y.-M.}\ \bibnamefont
  {He}}, \bibinfo {author} {\bibfnamefont {G.}~\bibnamefont {Clark}}, \bibinfo
  {author} {\bibfnamefont {J.~R.}\ \bibnamefont {Schaibley}}, \bibinfo {author}
  {\bibfnamefont {Y.}~\bibnamefont {He}}, \bibinfo {author} {\bibfnamefont
  {M.-C.}\ \bibnamefont {Chen}}, \bibinfo {author} {\bibfnamefont {Y.-J.}\
  \bibnamefont {Wei}}, \bibinfo {author} {\bibfnamefont {X.}~\bibnamefont
  {Ding}}, \bibinfo {author} {\bibfnamefont {Q.}~\bibnamefont {Zhang}},
  \bibinfo {author} {\bibfnamefont {W.}~\bibnamefont {Yao}}, \bibinfo {author}
  {\bibfnamefont {X.}~\bibnamefont {Xu}},\ and\ \bibinfo {author} {\bibnamefont
  {et~al.}},\ }\bibfield  {title} {\enquote {\bibinfo {title} {Single quantum
  emitters in monolayer semiconductors},}\ }\href
  {https://doi.org/10.1038/nnano.2015.75} {\bibfield  {journal} {\bibinfo
  {journal} {Nat. Nanotechnol.}\ }\textbf {\bibinfo {volume} {10}},\ \bibinfo
  {pages} {497–502} (\bibinfo {year} {2015})}\BibitemShut {NoStop}%
\bibitem [{\citenamefont {Tonndorf}\ \emph {et~al.}(2015)\citenamefont
  {Tonndorf}, \citenamefont {Schmidt}, \citenamefont {Schneider}, \citenamefont
  {Kern}, \citenamefont {Buscema}, \citenamefont {Steele}, \citenamefont
  {Castellanos-Gomez}, \citenamefont {van~der Zant}, \citenamefont
  {de~Vasconcellos},\ and\ \citenamefont {Bratschitsch}}]{Tonndorf15}%
  \BibitemOpen
  \bibfield  {author} {\bibinfo {author} {\bibfnamefont {P.}~\bibnamefont
  {Tonndorf}}, \bibinfo {author} {\bibfnamefont {R.}~\bibnamefont {Schmidt}},
  \bibinfo {author} {\bibfnamefont {R.}~\bibnamefont {Schneider}}, \bibinfo
  {author} {\bibfnamefont {J.}~\bibnamefont {Kern}}, \bibinfo {author}
  {\bibfnamefont {M.}~\bibnamefont {Buscema}}, \bibinfo {author} {\bibfnamefont
  {G.~A.}\ \bibnamefont {Steele}}, \bibinfo {author} {\bibfnamefont
  {A.}~\bibnamefont {Castellanos-Gomez}}, \bibinfo {author} {\bibfnamefont
  {H.~S.~J.}\ \bibnamefont {van~der Zant}}, \bibinfo {author} {\bibfnamefont
  {S.~M.}\ \bibnamefont {de~Vasconcellos}},\ and\ \bibinfo {author}
  {\bibfnamefont {R.}~\bibnamefont {Bratschitsch}},\ }\bibfield  {title}
  {\enquote {\bibinfo {title} {Single-photon emission from localized excitons
  in an atomically thin semiconductor},}\ }\href
  {https://doi.org/10.1364/OPTICA.2.000347} {\bibfield  {journal} {\bibinfo
  {journal} {Optica}\ }\textbf {\bibinfo {volume} {2}},\ \bibinfo {pages}
  {347--352} (\bibinfo {year} {2015})}\BibitemShut {NoStop}%
\bibitem [{\citenamefont {Palacios-Berraquero}\ \emph
  {et~al.}(2017)\citenamefont {Palacios-Berraquero}, \citenamefont {Kara},
  \citenamefont {Montblanch}, \citenamefont {Barbone}, \citenamefont
  {Latawiec}, \citenamefont {Yoon}, \citenamefont {Ott}, \citenamefont
  {Loncar}, \citenamefont {Ferrari},\ and\ \citenamefont
  {Atatüre}}]{Carmen2017}%
  \BibitemOpen
  \bibfield  {author} {\bibinfo {author} {\bibfnamefont {C.}~\bibnamefont
  {Palacios-Berraquero}}, \bibinfo {author} {\bibfnamefont {D.~M.}\
  \bibnamefont {Kara}}, \bibinfo {author} {\bibfnamefont {A.~R.-P.}\
  \bibnamefont {Montblanch}}, \bibinfo {author} {\bibfnamefont
  {M.}~\bibnamefont {Barbone}}, \bibinfo {author} {\bibfnamefont
  {P.}~\bibnamefont {Latawiec}}, \bibinfo {author} {\bibfnamefont
  {D.}~\bibnamefont {Yoon}}, \bibinfo {author} {\bibfnamefont {A.~K.}\
  \bibnamefont {Ott}}, \bibinfo {author} {\bibfnamefont {M.}~\bibnamefont
  {Loncar}}, \bibinfo {author} {\bibfnamefont {A.~C.}\ \bibnamefont
  {Ferrari}},\ and\ \bibinfo {author} {\bibfnamefont {M.}~\bibnamefont
  {Atatüre}},\ }\bibfield  {title} {\enquote {\bibinfo {title} {Large-scale
  quantum-emitter arrays in atomically thin semiconductors},}\ }\href
  {https://doi.org/10.1038/ncomms15093} {\bibfield  {journal} {\bibinfo
  {journal} {Nature Communications}\ }\textbf {\bibinfo {volume} {8}},\
  \bibinfo {pages} {15093} (\bibinfo {year} {2017})}\BibitemShut {NoStop}%
\bibitem [{\citenamefont {Branny}\ \emph {et~al.}(2017)\citenamefont {Branny},
  \citenamefont {Kumar}, \citenamefont {Proux},\ and\ \citenamefont
  {Gerardot}}]{Branny2017}%
  \BibitemOpen
  \bibfield  {author} {\bibinfo {author} {\bibfnamefont {A.}~\bibnamefont
  {Branny}}, \bibinfo {author} {\bibfnamefont {S.}~\bibnamefont {Kumar}},
  \bibinfo {author} {\bibfnamefont {R.}~\bibnamefont {Proux}},\ and\ \bibinfo
  {author} {\bibfnamefont {B.~D.}\ \bibnamefont {Gerardot}},\ }\bibfield
  {title} {\enquote {\bibinfo {title} {Deterministic strain-induced arrays of
  quantum emitters in a two-dimensional semiconductor},}\ }\href
  {https://doi.org/10.1038/ncomms15053} {\bibfield  {journal} {\bibinfo
  {journal} {Nature Communications}\ }\textbf {\bibinfo {volume} {8}},\
  \bibinfo {pages} {15053} (\bibinfo {year} {2017})}\BibitemShut {NoStop}%
\bibitem [{\citenamefont {Tran}\ \emph
  {et~al.}(2016{\natexlab{a}})\citenamefont {Tran}, \citenamefont {Bray},
  \citenamefont {Ford}, \citenamefont {Toth},\ and\ \citenamefont
  {Aharonovich}}]{tran_bray_ford_toth_aharonovich_2016}%
  \BibitemOpen
  \bibfield  {author} {\bibinfo {author} {\bibfnamefont {T.~T.}\ \bibnamefont
  {Tran}}, \bibinfo {author} {\bibfnamefont {K.}~\bibnamefont {Bray}}, \bibinfo
  {author} {\bibfnamefont {M.~J.}\ \bibnamefont {Ford}}, \bibinfo {author}
  {\bibfnamefont {M.}~\bibnamefont {Toth}},\ and\ \bibinfo {author}
  {\bibfnamefont {I.}~\bibnamefont {Aharonovich}},\ }\bibfield  {title}
  {\enquote {\bibinfo {title} {Quantum emission from hexagonal boron nitride
  monolayers},}\ }\href {https://doi.org/10.1038/nnano.2015.242} {\bibfield
  {journal} {\bibinfo  {journal} {Nat. Nanotechnol.}\ }\textbf {\bibinfo
  {volume} {11}},\ \bibinfo {pages} {37–41} (\bibinfo {year}
  {2016}{\natexlab{a}})}\BibitemShut {NoStop}%
\bibitem [{\citenamefont {Nikolay}\ \emph {et~al.}(2019)\citenamefont
  {Nikolay}, \citenamefont {Mendelson}, \citenamefont {\"{O}zelci},
  \citenamefont {Sontheimer}, \citenamefont {B\"{o}hm}, \citenamefont {Kewes},
  \citenamefont {Toth}, \citenamefont {Aharonovich},\ and\ \citenamefont
  {Benson}}]{Nikolay:19}%
  \BibitemOpen
  \bibfield  {author} {\bibinfo {author} {\bibfnamefont {N.}~\bibnamefont
  {Nikolay}}, \bibinfo {author} {\bibfnamefont {N.}~\bibnamefont {Mendelson}},
  \bibinfo {author} {\bibfnamefont {E.}~\bibnamefont {\"{O}zelci}}, \bibinfo
  {author} {\bibfnamefont {B.}~\bibnamefont {Sontheimer}}, \bibinfo {author}
  {\bibfnamefont {F.}~\bibnamefont {B\"{o}hm}}, \bibinfo {author}
  {\bibfnamefont {G.}~\bibnamefont {Kewes}}, \bibinfo {author} {\bibfnamefont
  {M.}~\bibnamefont {Toth}}, \bibinfo {author} {\bibfnamefont {I.}~\bibnamefont
  {Aharonovich}},\ and\ \bibinfo {author} {\bibfnamefont {O.}~\bibnamefont
  {Benson}},\ }\bibfield  {title} {\enquote {\bibinfo {title} {Direct
  measurement of quantum efficiency of single-photon emitters in hexagonal
  boron nitride},}\ }\href {https://doi.org/10.1364/OPTICA.6.001084} {\bibfield
   {journal} {\bibinfo  {journal} {Optica}\ }\textbf {\bibinfo {volume} {6}},\
  \bibinfo {pages} {1084--1088} (\bibinfo {year} {2019})}\BibitemShut {NoStop}%
\bibitem [{\citenamefont {Vogl}\ \emph {et~al.}(2018)\citenamefont {Vogl},
  \citenamefont {Campbell}, \citenamefont {Buchler}, \citenamefont {Lu},\ and\
  \citenamefont {Lam}}]{vogl_campbell_buchler_lu_lam_2018}%
  \BibitemOpen
  \bibfield  {author} {\bibinfo {author} {\bibfnamefont {T.}~\bibnamefont
  {Vogl}}, \bibinfo {author} {\bibfnamefont {G.}~\bibnamefont {Campbell}},
  \bibinfo {author} {\bibfnamefont {B.~C.}\ \bibnamefont {Buchler}}, \bibinfo
  {author} {\bibfnamefont {Y.}~\bibnamefont {Lu}},\ and\ \bibinfo {author}
  {\bibfnamefont {P.~K.}\ \bibnamefont {Lam}},\ }\bibfield  {title} {\enquote
  {\bibinfo {title} {Fabrication and deterministic transfer of high-quality
  quantum emitters in hexagonal boron nitride},}\ }\href
  {https://doi.org/10.1021/acsphotonics.8b00127} {\bibfield  {journal}
  {\bibinfo  {journal} {ACS Photonics}\ }\textbf {\bibinfo {volume} {5}},\
  \bibinfo {pages} {2305–2312} (\bibinfo {year} {2018})}\BibitemShut
  {NoStop}%
\bibitem [{\citenamefont {Kubanek}(2022)}]{kubanek_2022}%
  \BibitemOpen
  \bibfield  {author} {\bibinfo {author} {\bibfnamefont {A.}~\bibnamefont
  {Kubanek}},\ }\bibfield  {title} {\enquote {\bibinfo {title} {Coherent
  quantum emitters in hexagonal boron nitride},}\ }\href
  {https://doi.org/10.1002/qute.202200009} {\bibfield  {journal} {\bibinfo
  {journal} {Adv. Quantum Technol.}\ ,\ \bibinfo {pages} {2200009}} (\bibinfo
  {year} {2022})}\BibitemShut {NoStop}%
\bibitem [{\citenamefont {Chejanovsky}\ \emph {et~al.}(2016)\citenamefont
  {Chejanovsky}, \citenamefont {Rezai}, \citenamefont {Paolucci}, \citenamefont
  {Kim}, \citenamefont {Rendler}, \citenamefont {Rouabeh}, \citenamefont
  {F{\'a}varo~de Oliveira}, \citenamefont {Herlinger}, \citenamefont
  {Denisenko}, \citenamefont {Yang}, \citenamefont {Gerhardt}, \citenamefont
  {Finkler}, \citenamefont {Smet},\ and\ \citenamefont
  {Wrachtrup}}]{10.1021/acs.nanolett.6b03268}%
  \BibitemOpen
  \bibfield  {author} {\bibinfo {author} {\bibfnamefont {N.}~\bibnamefont
  {Chejanovsky}}, \bibinfo {author} {\bibfnamefont {M.}~\bibnamefont {Rezai}},
  \bibinfo {author} {\bibfnamefont {F.}~\bibnamefont {Paolucci}}, \bibinfo
  {author} {\bibfnamefont {Y.}~\bibnamefont {Kim}}, \bibinfo {author}
  {\bibfnamefont {T.}~\bibnamefont {Rendler}}, \bibinfo {author} {\bibfnamefont
  {W.}~\bibnamefont {Rouabeh}}, \bibinfo {author} {\bibfnamefont
  {F.}~\bibnamefont {F{\'a}varo~de Oliveira}}, \bibinfo {author} {\bibfnamefont
  {P.}~\bibnamefont {Herlinger}}, \bibinfo {author} {\bibfnamefont
  {A.}~\bibnamefont {Denisenko}}, \bibinfo {author} {\bibfnamefont
  {S.}~\bibnamefont {Yang}}, \bibinfo {author} {\bibfnamefont {I.}~\bibnamefont
  {Gerhardt}}, \bibinfo {author} {\bibfnamefont {A.}~\bibnamefont {Finkler}},
  \bibinfo {author} {\bibfnamefont {J.~H.}\ \bibnamefont {Smet}},\ and\
  \bibinfo {author} {\bibfnamefont {J.}~\bibnamefont {Wrachtrup}},\ }\bibfield
  {title} {\enquote {\bibinfo {title} {Structural attributes and photodynamics
  of visible spectrum quantum emitters in hexagonal boron nitride},}\ }\href
  {https://doi.org/10.1021/acs.nanolett.6b03268} {\bibfield  {journal}
  {\bibinfo  {journal} {Nano Lett.}\ }\textbf {\bibinfo {volume} {16}},\
  \bibinfo {pages} {7037--7045} (\bibinfo {year} {2016})}\BibitemShut {NoStop}%
\bibitem [{\citenamefont {Xu}\ \emph {et~al.}(2021)\citenamefont {Xu},
  \citenamefont {Martin}, \citenamefont {Sychev}, \citenamefont {Lagutchev},
  \citenamefont {Chen}, \citenamefont {Taniguchi}, \citenamefont {Watanabe},
  \citenamefont {Shalaev},\ and\ \citenamefont
  {Boltasseva}}]{xu_martin_sychev_lagutchev_chen_taniguchi_watanabe_shalaev_boltasseva_2021}%
  \BibitemOpen
  \bibfield  {author} {\bibinfo {author} {\bibfnamefont {X.}~\bibnamefont
  {Xu}}, \bibinfo {author} {\bibfnamefont {Z.~O.}\ \bibnamefont {Martin}},
  \bibinfo {author} {\bibfnamefont {D.}~\bibnamefont {Sychev}}, \bibinfo
  {author} {\bibfnamefont {A.~S.}\ \bibnamefont {Lagutchev}}, \bibinfo {author}
  {\bibfnamefont {Y.~P.}\ \bibnamefont {Chen}}, \bibinfo {author}
  {\bibfnamefont {T.}~\bibnamefont {Taniguchi}}, \bibinfo {author}
  {\bibfnamefont {K.}~\bibnamefont {Watanabe}}, \bibinfo {author}
  {\bibfnamefont {V.~M.}\ \bibnamefont {Shalaev}},\ and\ \bibinfo {author}
  {\bibfnamefont {A.}~\bibnamefont {Boltasseva}},\ }\bibfield  {title}
  {\enquote {\bibinfo {title} {Creating quantum emitters in hexagonal boron
  nitride deterministically on chip-compatible substrates},}\ }\href
  {https://doi.org/10.1021/acs.nanolett.1c02640} {\bibfield  {journal}
  {\bibinfo  {journal} {Nano Lett.}\ }\textbf {\bibinfo {volume} {21}},\
  \bibinfo {pages} {8182–8189} (\bibinfo {year} {2021})}\BibitemShut
  {NoStop}%
\bibitem [{\citenamefont {Proscia}\ \emph {et~al.}(2018)\citenamefont
  {Proscia}, \citenamefont {Shotan}, \citenamefont {Jayakumar}, \citenamefont
  {Reddy}, \citenamefont {Cohen}, \citenamefont {Dollar}, \citenamefont
  {Alkauskas}, \citenamefont {Doherty}, \citenamefont {Meriles},\ and\
  \citenamefont {Menon}}]{10.1364/OPTICA.5.001128}%
  \BibitemOpen
  \bibfield  {author} {\bibinfo {author} {\bibfnamefont {N.~V.}\ \bibnamefont
  {Proscia}}, \bibinfo {author} {\bibfnamefont {Z.}~\bibnamefont {Shotan}},
  \bibinfo {author} {\bibfnamefont {H.}~\bibnamefont {Jayakumar}}, \bibinfo
  {author} {\bibfnamefont {P.}~\bibnamefont {Reddy}}, \bibinfo {author}
  {\bibfnamefont {C.}~\bibnamefont {Cohen}}, \bibinfo {author} {\bibfnamefont
  {M.}~\bibnamefont {Dollar}}, \bibinfo {author} {\bibfnamefont
  {A.}~\bibnamefont {Alkauskas}}, \bibinfo {author} {\bibfnamefont
  {M.}~\bibnamefont {Doherty}}, \bibinfo {author} {\bibfnamefont {C.~A.}\
  \bibnamefont {Meriles}},\ and\ \bibinfo {author} {\bibfnamefont {V.~M.}\
  \bibnamefont {Menon}},\ }\bibfield  {title} {\enquote {\bibinfo {title}
  {Near-deterministic activation of room-temperature quantum emitters in
  hexagonal boron nitride},}\ }\href {https://doi.org/10.1364/OPTICA.5.001128}
  {\bibfield  {journal} {\bibinfo  {journal} {Optica}\ }\textbf {\bibinfo
  {volume} {5}},\ \bibinfo {pages} {1128--1134} (\bibinfo {year}
  {2018})}\BibitemShut {NoStop}%
\bibitem [{\citenamefont {Li}\ \emph {et~al.}(2021)\citenamefont {Li},
  \citenamefont {Mendelson}, \citenamefont {Ritika}, \citenamefont {Chen},
  \citenamefont {Xu}, \citenamefont {Toth},\ and\ \citenamefont
  {Aharonovich}}]{doi:10.1021/acs.nanolett.1c00685}%
  \BibitemOpen
  \bibfield  {author} {\bibinfo {author} {\bibfnamefont {C.}~\bibnamefont
  {Li}}, \bibinfo {author} {\bibfnamefont {N.}~\bibnamefont {Mendelson}},
  \bibinfo {author} {\bibfnamefont {R.}~\bibnamefont {Ritika}}, \bibinfo
  {author} {\bibfnamefont {Y.}~\bibnamefont {Chen}}, \bibinfo {author}
  {\bibfnamefont {Z.-Q.}\ \bibnamefont {Xu}}, \bibinfo {author} {\bibfnamefont
  {M.}~\bibnamefont {Toth}},\ and\ \bibinfo {author} {\bibfnamefont
  {I.}~\bibnamefont {Aharonovich}},\ }\bibfield  {title} {\enquote {\bibinfo
  {title} {Scalable and deterministic fabrication of quantum emitter arrays
  from hexagonal boron nitride},}\ }\href
  {https://doi.org/10.1021/acs.nanolett.1c00685} {\bibfield  {journal}
  {\bibinfo  {journal} {Nano Lett.}\ }\textbf {\bibinfo {volume} {21}},\
  \bibinfo {pages} {3626--3632} (\bibinfo {year} {2021})},\ \bibinfo {note}
  {pMID: 33870699}\BibitemShut {NoStop}%
\bibitem [{\citenamefont {Hou}\ \emph {et~al.}(2017)\citenamefont {Hou},
  \citenamefont {Birowosuto}, \citenamefont {Umar}, \citenamefont {Anicet},
  \citenamefont {Tay}, \citenamefont {Coquet}, \citenamefont {Tay},
  \citenamefont {Wang},\ and\ \citenamefont {Teo}}]{Hou_2017}%
  \BibitemOpen
  \bibfield  {author} {\bibinfo {author} {\bibfnamefont {S.}~\bibnamefont
  {Hou}}, \bibinfo {author} {\bibfnamefont {M.~D.}\ \bibnamefont {Birowosuto}},
  \bibinfo {author} {\bibfnamefont {S.}~\bibnamefont {Umar}}, \bibinfo {author}
  {\bibfnamefont {M.~A.}\ \bibnamefont {Anicet}}, \bibinfo {author}
  {\bibfnamefont {R.~Y.}\ \bibnamefont {Tay}}, \bibinfo {author} {\bibfnamefont
  {P.}~\bibnamefont {Coquet}}, \bibinfo {author} {\bibfnamefont {B.~K.}\
  \bibnamefont {Tay}}, \bibinfo {author} {\bibfnamefont {H.}~\bibnamefont
  {Wang}},\ and\ \bibinfo {author} {\bibfnamefont {E.~H.~T.}\ \bibnamefont
  {Teo}},\ }\bibfield  {title} {\enquote {\bibinfo {title} {Localized emission
  from laser-irradiated defects in 2d hexagonal boron nitride},}\ }\href
  {https://doi.org/10.1088/2053-1583/aa8e61} {\bibfield  {journal} {\bibinfo
  {journal} {2D Mater.}\ }\textbf {\bibinfo {volume} {5}},\ \bibinfo {pages}
  {015010} (\bibinfo {year} {2017})}\BibitemShut {NoStop}%
\bibitem [{\citenamefont {Vogl}\ \emph
  {et~al.}(2019{\natexlab{b}})\citenamefont {Vogl}, \citenamefont {Sripathy},
  \citenamefont {Sharma}, \citenamefont {Reddy}, \citenamefont {Sullivan},
  \citenamefont {Machacek}, \citenamefont {Zhang}, \citenamefont {Karouta},
  \citenamefont {Buchler}, \citenamefont {Doherty},\ and\ \citenamefont
  {et~al.}}]{vogl_sripathy_sharma_reddy_sullivan_machacek_zhang_karouta_buchler_doherty}%
  \BibitemOpen
  \bibfield  {author} {\bibinfo {author} {\bibfnamefont {T.}~\bibnamefont
  {Vogl}}, \bibinfo {author} {\bibfnamefont {K.}~\bibnamefont {Sripathy}},
  \bibinfo {author} {\bibfnamefont {A.}~\bibnamefont {Sharma}}, \bibinfo
  {author} {\bibfnamefont {P.}~\bibnamefont {Reddy}}, \bibinfo {author}
  {\bibfnamefont {J.}~\bibnamefont {Sullivan}}, \bibinfo {author}
  {\bibfnamefont {J.~R.}\ \bibnamefont {Machacek}}, \bibinfo {author}
  {\bibfnamefont {L.}~\bibnamefont {Zhang}}, \bibinfo {author} {\bibfnamefont
  {F.}~\bibnamefont {Karouta}}, \bibinfo {author} {\bibfnamefont {B.~C.}\
  \bibnamefont {Buchler}}, \bibinfo {author} {\bibfnamefont {M.~W.}\
  \bibnamefont {Doherty}},\ and\ \bibinfo {author} {\bibnamefont {et~al.}},\
  }\bibfield  {title} {\enquote {\bibinfo {title} {Radiation tolerance of
  two-dimensional material-based devices for space applications},}\ }\href
  {https://doi.org/10.1038/s41467-019-09219-5} {\bibfield  {journal} {\bibinfo
  {journal} {Nat. Commun.}\ }\textbf {\bibinfo {volume} {10}},\ \bibinfo
  {pages} {1202} (\bibinfo {year} {2019}{\natexlab{b}})}\BibitemShut {NoStop}%
\bibitem [{\citenamefont {Choi}\ \emph {et~al.}(2016)\citenamefont {Choi},
  \citenamefont {Tran}, \citenamefont {Elbadawi}, \citenamefont {Lobo},
  \citenamefont {Wang}, \citenamefont {Juodkazis}, \citenamefont {Seniutinas},
  \citenamefont {Toth},\ and\ \citenamefont
  {Aharonovich}}]{10.1021/acsami.6b09875}%
  \BibitemOpen
  \bibfield  {author} {\bibinfo {author} {\bibfnamefont {S.}~\bibnamefont
  {Choi}}, \bibinfo {author} {\bibfnamefont {T.~T.}\ \bibnamefont {Tran}},
  \bibinfo {author} {\bibfnamefont {C.}~\bibnamefont {Elbadawi}}, \bibinfo
  {author} {\bibfnamefont {C.}~\bibnamefont {Lobo}}, \bibinfo {author}
  {\bibfnamefont {X.}~\bibnamefont {Wang}}, \bibinfo {author} {\bibfnamefont
  {S.}~\bibnamefont {Juodkazis}}, \bibinfo {author} {\bibfnamefont
  {G.}~\bibnamefont {Seniutinas}}, \bibinfo {author} {\bibfnamefont
  {M.}~\bibnamefont {Toth}},\ and\ \bibinfo {author} {\bibfnamefont
  {I.}~\bibnamefont {Aharonovich}},\ }\bibfield  {title} {\enquote {\bibinfo
  {title} {Engineering and localization of quantum emitters in large hexagonal
  boron nitride layers},}\ }\href {https://doi.org/10.1021/acsami.6b09875}
  {\bibfield  {journal} {\bibinfo  {journal} {ACS Appl. Mater. Interfaces}\
  }\textbf {\bibinfo {volume} {8}},\ \bibinfo {pages} {29642--29648} (\bibinfo
  {year} {2016})}\BibitemShut {NoStop}%
\bibitem [{\citenamefont {Fournier}\ \emph {et~al.}(2021)\citenamefont
  {Fournier}, \citenamefont {Plaud}, \citenamefont {Roux}, \citenamefont
  {Pierret}, \citenamefont {Rosticher}, \citenamefont {Watanabe}, \citenamefont
  {Taniguchi}, \citenamefont {Buil}, \citenamefont {Quélin}, \citenamefont
  {Barjon},\ and\ \citenamefont {et~al.}}]{naturecomm}%
  \BibitemOpen
  \bibfield  {author} {\bibinfo {author} {\bibfnamefont {C.}~\bibnamefont
  {Fournier}}, \bibinfo {author} {\bibfnamefont {A.}~\bibnamefont {Plaud}},
  \bibinfo {author} {\bibfnamefont {S.}~\bibnamefont {Roux}}, \bibinfo {author}
  {\bibfnamefont {A.}~\bibnamefont {Pierret}}, \bibinfo {author} {\bibfnamefont
  {M.}~\bibnamefont {Rosticher}}, \bibinfo {author} {\bibfnamefont
  {K.}~\bibnamefont {Watanabe}}, \bibinfo {author} {\bibfnamefont
  {T.}~\bibnamefont {Taniguchi}}, \bibinfo {author} {\bibfnamefont
  {S.}~\bibnamefont {Buil}}, \bibinfo {author} {\bibfnamefont {X.}~\bibnamefont
  {Quélin}}, \bibinfo {author} {\bibfnamefont {J.}~\bibnamefont {Barjon}},\
  and\ \bibinfo {author} {\bibnamefont {et~al.}},\ }\bibfield  {title}
  {\enquote {\bibinfo {title} {Position-controlled quantum emitters with
  reproducible emission wavelength in hexagonal boron nitride},}\ }\href
  {https://doi.org/10.1038/s41467-021-24019-6} {\bibfield  {journal} {\bibinfo
  {journal} {Nat. Commun.}\ }\textbf {\bibinfo {volume} {12}},\ \bibinfo
  {pages} {3779} (\bibinfo {year} {2021})}\BibitemShut {NoStop}%
\bibitem [{\citenamefont {Gale}\ \emph {et~al.}(2022)\citenamefont {Gale},
  \citenamefont {Li}, \citenamefont {Chen}, \citenamefont {Watanabe},
  \citenamefont {Taniguchi}, \citenamefont {Aharonovich},\ and\ \citenamefont
  {Toth}}]{doi:10.1021/acsphotonics.2c00631}%
  \BibitemOpen
  \bibfield  {author} {\bibinfo {author} {\bibfnamefont {A.}~\bibnamefont
  {Gale}}, \bibinfo {author} {\bibfnamefont {C.}~\bibnamefont {Li}}, \bibinfo
  {author} {\bibfnamefont {Y.}~\bibnamefont {Chen}}, \bibinfo {author}
  {\bibfnamefont {K.}~\bibnamefont {Watanabe}}, \bibinfo {author}
  {\bibfnamefont {T.}~\bibnamefont {Taniguchi}}, \bibinfo {author}
  {\bibfnamefont {I.}~\bibnamefont {Aharonovich}},\ and\ \bibinfo {author}
  {\bibfnamefont {M.}~\bibnamefont {Toth}},\ }\bibfield  {title} {\enquote
  {\bibinfo {title} {Site-specific fabrication of blue quantum emitters in
  hexagonal boron nitride},}\ }\href
  {https://doi.org/10.1021/acsphotonics.2c00631} {\bibfield  {journal}
  {\bibinfo  {journal} {ACS Photonics}\ }\textbf {\bibinfo {volume} {9}},\
  \bibinfo {pages} {2170–2177} (\bibinfo {year} {2022})}\BibitemShut
  {NoStop}%
\bibitem [{\citenamefont {Shevitski}\ \emph {et~al.}(2019)\citenamefont
  {Shevitski}, \citenamefont {Gilbert}, \citenamefont {Chen}, \citenamefont
  {Kastl}, \citenamefont {Barnard}, \citenamefont {Wong}, \citenamefont
  {Ogletree}, \citenamefont {Watanabe}, \citenamefont {Taniguchi},
  \citenamefont {Zettl},\ and\ \citenamefont {Aloni}}]{PhysRevB.100.155419}%
  \BibitemOpen
  \bibfield  {author} {\bibinfo {author} {\bibfnamefont {B.}~\bibnamefont
  {Shevitski}}, \bibinfo {author} {\bibfnamefont {S.~M.}\ \bibnamefont
  {Gilbert}}, \bibinfo {author} {\bibfnamefont {C.~T.}\ \bibnamefont {Chen}},
  \bibinfo {author} {\bibfnamefont {C.}~\bibnamefont {Kastl}}, \bibinfo
  {author} {\bibfnamefont {E.~S.}\ \bibnamefont {Barnard}}, \bibinfo {author}
  {\bibfnamefont {E.}~\bibnamefont {Wong}}, \bibinfo {author} {\bibfnamefont
  {D.~F.}\ \bibnamefont {Ogletree}}, \bibinfo {author} {\bibfnamefont
  {K.}~\bibnamefont {Watanabe}}, \bibinfo {author} {\bibfnamefont
  {T.}~\bibnamefont {Taniguchi}}, \bibinfo {author} {\bibfnamefont
  {A.}~\bibnamefont {Zettl}},\ and\ \bibinfo {author} {\bibfnamefont
  {S.}~\bibnamefont {Aloni}},\ }\bibfield  {title} {\enquote {\bibinfo {title}
  {Blue-light-emitting color centers in high-quality hexagonal boron
  nitride},}\ }\href {https://doi.org/10.1103/PhysRevB.100.155419} {\bibfield
  {journal} {\bibinfo  {journal} {Phys. Rev. B}\ }\textbf {\bibinfo {volume}
  {100}},\ \bibinfo {pages} {155419} (\bibinfo {year} {2019})}\BibitemShut
  {NoStop}%
\bibitem [{\citenamefont {Ngoc My~Duong}\ \emph {et~al.}(2018)\citenamefont
  {Ngoc My~Duong}, \citenamefont {Nguyen}, \citenamefont {Kianinia},
  \citenamefont {Ohshima}, \citenamefont {Abe}, \citenamefont {Watanabe},
  \citenamefont {Taniguchi}, \citenamefont {Edgar}, \citenamefont
  {Aharonovich}, \citenamefont {Toth},\ and\ \citenamefont {et~al.}}]{ngoc}%
  \BibitemOpen
  \bibfield  {author} {\bibinfo {author} {\bibfnamefont {H.}~\bibnamefont {Ngoc
  My~Duong}}, \bibinfo {author} {\bibfnamefont {M.~A.~P.}\ \bibnamefont
  {Nguyen}}, \bibinfo {author} {\bibfnamefont {M.}~\bibnamefont {Kianinia}},
  \bibinfo {author} {\bibfnamefont {T.}~\bibnamefont {Ohshima}}, \bibinfo
  {author} {\bibfnamefont {H.}~\bibnamefont {Abe}}, \bibinfo {author}
  {\bibfnamefont {K.}~\bibnamefont {Watanabe}}, \bibinfo {author}
  {\bibfnamefont {T.}~\bibnamefont {Taniguchi}}, \bibinfo {author}
  {\bibfnamefont {J.~H.}\ \bibnamefont {Edgar}}, \bibinfo {author}
  {\bibfnamefont {I.}~\bibnamefont {Aharonovich}}, \bibinfo {author}
  {\bibfnamefont {M.}~\bibnamefont {Toth}},\ and\ \bibinfo {author}
  {\bibnamefont {et~al.}},\ }\bibfield  {title} {\enquote {\bibinfo {title}
  {Effects of high-energy electron irradiation on quantum emitters in hexagonal
  boron nitride},}\ }\href {https://doi.org/10.1021/acsami.8b07506} {\bibfield
  {journal} {\bibinfo  {journal} {ACS Appl. Mater. Interfaces}\ }\textbf
  {\bibinfo {volume} {10}},\ \bibinfo {pages} {24886–24891} (\bibinfo {year}
  {2018})}\BibitemShut {NoStop}%
\bibitem [{\citenamefont {Bourrellier}\ \emph {et~al.}(2016)\citenamefont
  {Bourrellier}, \citenamefont {Meuret}, \citenamefont {Tararan}, \citenamefont
  {Stéphan}, \citenamefont {Kociak}, \citenamefont {Tizei},\ and\
  \citenamefont {Zobelli}}]{doi:10.1021/acs.nanolett.6b01368}%
  \BibitemOpen
  \bibfield  {author} {\bibinfo {author} {\bibfnamefont {R.}~\bibnamefont
  {Bourrellier}}, \bibinfo {author} {\bibfnamefont {S.}~\bibnamefont {Meuret}},
  \bibinfo {author} {\bibfnamefont {A.}~\bibnamefont {Tararan}}, \bibinfo
  {author} {\bibfnamefont {O.}~\bibnamefont {Stéphan}}, \bibinfo {author}
  {\bibfnamefont {M.}~\bibnamefont {Kociak}}, \bibinfo {author} {\bibfnamefont
  {L.~H.~G.}\ \bibnamefont {Tizei}},\ and\ \bibinfo {author} {\bibfnamefont
  {A.}~\bibnamefont {Zobelli}},\ }\bibfield  {title} {\enquote {\bibinfo
  {title} {Bright uv single photon emission at point defects in {h-BN}},}\
  }\href {https://doi.org/10.1021/acs.nanolett.6b01368} {\bibfield  {journal}
  {\bibinfo  {journal} {Nano Lett.}\ }\textbf {\bibinfo {volume} {16}},\
  \bibinfo {pages} {4317--4321} (\bibinfo {year} {2016})}\BibitemShut {NoStop}%
\bibitem [{\citenamefont {Tran}\ \emph
  {et~al.}(2016{\natexlab{b}})\citenamefont {Tran}, \citenamefont {Elbadawi},
  \citenamefont {Totonjian}, \citenamefont {Lobo}, \citenamefont {Grosso},
  \citenamefont {Moon}, \citenamefont {Englund}, \citenamefont {Ford},
  \citenamefont {Aharonovich}, \citenamefont {Toth},\ and\ \citenamefont
  {et~al.}}]{tran_elbadawi_totonjian_lobo_grosso_moon_englund_ford_aharonovich_toth_et}%
  \BibitemOpen
  \bibfield  {author} {\bibinfo {author} {\bibfnamefont {T.~T.}\ \bibnamefont
  {Tran}}, \bibinfo {author} {\bibfnamefont {C.}~\bibnamefont {Elbadawi}},
  \bibinfo {author} {\bibfnamefont {D.}~\bibnamefont {Totonjian}}, \bibinfo
  {author} {\bibfnamefont {C.~J.}\ \bibnamefont {Lobo}}, \bibinfo {author}
  {\bibfnamefont {G.}~\bibnamefont {Grosso}}, \bibinfo {author} {\bibfnamefont
  {H.}~\bibnamefont {Moon}}, \bibinfo {author} {\bibfnamefont {D.~R.}\
  \bibnamefont {Englund}}, \bibinfo {author} {\bibfnamefont {M.~J.}\
  \bibnamefont {Ford}}, \bibinfo {author} {\bibfnamefont {I.}~\bibnamefont
  {Aharonovich}}, \bibinfo {author} {\bibfnamefont {M.}~\bibnamefont {Toth}},\
  and\ \bibinfo {author} {\bibnamefont {et~al.}},\ }\bibfield  {title}
  {\enquote {\bibinfo {title} {Robust multicolor single photon emission from
  point defects in hexagonal boron nitride},}\ }\href
  {https://doi.org/10.1021/acsnano.6b03602} {\bibfield  {journal} {\bibinfo
  {journal} {ACS Nano}\ }\textbf {\bibinfo {volume} {10}},\ \bibinfo {pages}
  {7331–7338} (\bibinfo {year} {2016}{\natexlab{b}})}\BibitemShut {NoStop}%
\bibitem [{\citenamefont {Camphausen}\ \emph {et~al.}(2020)\citenamefont
  {Camphausen}, \citenamefont {Marini}, \citenamefont {Tawfik}, \citenamefont
  {Tran}, \citenamefont {Ford},\ and\ \citenamefont
  {Palomba}}]{doi:10.1063/5.0008242}%
  \BibitemOpen
  \bibfield  {author} {\bibinfo {author} {\bibfnamefont {R.}~\bibnamefont
  {Camphausen}}, \bibinfo {author} {\bibfnamefont {L.}~\bibnamefont {Marini}},
  \bibinfo {author} {\bibfnamefont {S.~A.}\ \bibnamefont {Tawfik}}, \bibinfo
  {author} {\bibfnamefont {T.~T.}\ \bibnamefont {Tran}}, \bibinfo {author}
  {\bibfnamefont {M.~J.}\ \bibnamefont {Ford}},\ and\ \bibinfo {author}
  {\bibfnamefont {S.}~\bibnamefont {Palomba}},\ }\bibfield  {title} {\enquote
  {\bibinfo {title} {Observation of near-infrared sub-poissonian photon
  emission in hexagonal boron nitride at room temperature},}\ }\href
  {https://doi.org/10.1063/5.0008242} {\bibfield  {journal} {\bibinfo
  {journal} {APL Photonics}\ }\textbf {\bibinfo {volume} {5}},\ \bibinfo
  {pages} {076103} (\bibinfo {year} {2020})}\BibitemShut {NoStop}%
\bibitem [{\citenamefont {Shaik}\ and\ \citenamefont
  {Palla}(2022)}]{Shaik2022}%
  \BibitemOpen
  \bibfield  {author} {\bibinfo {author} {\bibfnamefont {A.~B. D.-a.}\
  \bibnamefont {Shaik}}\ and\ \bibinfo {author} {\bibfnamefont
  {P.}~\bibnamefont {Palla}},\ }\bibfield  {title} {\enquote {\bibinfo {title}
  {Strain tunable quantum emission from atomic defects in hexagonal boron
  nitride for telecom-bands},}\ }\href
  {https://doi.org/10.1038/s41598-022-26061-w} {\bibfield  {journal} {\bibinfo
  {journal} {Sci. Rep.}\ }\textbf {\bibinfo {volume} {12}},\ \bibinfo {pages}
  {21673} (\bibinfo {year} {2022})}\BibitemShut {NoStop}%
\bibitem [{\citenamefont {Cholsuk}, \citenamefont {Suwanna},\ and\
  \citenamefont {Vogl}(2022)}]{Cholsuk2022}%
  \BibitemOpen
  \bibfield  {author} {\bibinfo {author} {\bibfnamefont {C.}~\bibnamefont
  {Cholsuk}}, \bibinfo {author} {\bibfnamefont {S.}~\bibnamefont {Suwanna}},\
  and\ \bibinfo {author} {\bibfnamefont {T.}~\bibnamefont {Vogl}},\ }\bibfield
  {title} {\enquote {\bibinfo {title} {Tailoring the emission wavelength of
  color centers in hexagonal boron nitride for quantum applications},}\ }\href
  {https://doi.org/10.3390/nano12142427} {\bibfield  {journal} {\bibinfo
  {journal} {Nanomaterials}\ }\textbf {\bibinfo {volume} {12}},\ \bibinfo
  {pages} {2427} (\bibinfo {year} {2022})}\BibitemShut {NoStop}%
\bibitem [{\citenamefont {Mendelson}\ \emph {et~al.}(2021)\citenamefont
  {Mendelson}, \citenamefont {Chugh}, \citenamefont {Reimers}, \citenamefont
  {Cheng}, \citenamefont {Gottscholl}, \citenamefont {Long}, \citenamefont
  {Mellor}, \citenamefont {Zettl}, \citenamefont {Dyakonov}, \citenamefont
  {Beton}, \citenamefont {Novikov}, \citenamefont {Jagadish}, \citenamefont
  {Tan}, \citenamefont {Ford}, \citenamefont {Toth}, \citenamefont {Bradac},\
  and\ \citenamefont {Aharonovich}}]{Mendelson2021}%
  \BibitemOpen
  \bibfield  {author} {\bibinfo {author} {\bibfnamefont {N.}~\bibnamefont
  {Mendelson}}, \bibinfo {author} {\bibfnamefont {D.}~\bibnamefont {Chugh}},
  \bibinfo {author} {\bibfnamefont {J.~R.}\ \bibnamefont {Reimers}}, \bibinfo
  {author} {\bibfnamefont {T.~S.}\ \bibnamefont {Cheng}}, \bibinfo {author}
  {\bibfnamefont {A.}~\bibnamefont {Gottscholl}}, \bibinfo {author}
  {\bibfnamefont {H.}~\bibnamefont {Long}}, \bibinfo {author} {\bibfnamefont
  {C.~J.}\ \bibnamefont {Mellor}}, \bibinfo {author} {\bibfnamefont
  {A.}~\bibnamefont {Zettl}}, \bibinfo {author} {\bibfnamefont
  {V.}~\bibnamefont {Dyakonov}}, \bibinfo {author} {\bibfnamefont {P.~H.}\
  \bibnamefont {Beton}}, \bibinfo {author} {\bibfnamefont {S.~V.}\ \bibnamefont
  {Novikov}}, \bibinfo {author} {\bibfnamefont {C.}~\bibnamefont {Jagadish}},
  \bibinfo {author} {\bibfnamefont {H.~H.}\ \bibnamefont {Tan}}, \bibinfo
  {author} {\bibfnamefont {M.~J.}\ \bibnamefont {Ford}}, \bibinfo {author}
  {\bibfnamefont {M.}~\bibnamefont {Toth}}, \bibinfo {author} {\bibfnamefont
  {C.}~\bibnamefont {Bradac}},\ and\ \bibinfo {author} {\bibfnamefont
  {I.}~\bibnamefont {Aharonovich}},\ }\bibfield  {title} {\enquote {\bibinfo
  {title} {Identifying carbon as the source of visible single-photon emission
  from hexagonal boron nitride},}\ }\href
  {https://doi.org/10.1038/s41563-020-00850-y} {\bibfield  {journal} {\bibinfo
  {journal} {Nat. Mater.}\ }\textbf {\bibinfo {volume} {20}},\ \bibinfo {pages}
  {321--328} (\bibinfo {year} {2021})}\BibitemShut {NoStop}%
\bibitem [{\citenamefont {Fischer}\ \emph {et~al.}(2021)\citenamefont
  {Fischer}, \citenamefont {Caridad}, \citenamefont {Sajid}, \citenamefont
  {Ghaderzadeh}, \citenamefont {Ghorbani-Asl}, \citenamefont {Gammelgaard},
  \citenamefont {Bøggild}, \citenamefont {Thygesen}, \citenamefont
  {Krasheninnikov}, \citenamefont {Xiao}, \citenamefont {Wubs},\ and\
  \citenamefont {Stenger}}]{doi:10.1126/sciadv.abe7138}%
  \BibitemOpen
  \bibfield  {author} {\bibinfo {author} {\bibfnamefont {M.}~\bibnamefont
  {Fischer}}, \bibinfo {author} {\bibfnamefont {J.~M.}\ \bibnamefont
  {Caridad}}, \bibinfo {author} {\bibfnamefont {A.}~\bibnamefont {Sajid}},
  \bibinfo {author} {\bibfnamefont {S.}~\bibnamefont {Ghaderzadeh}}, \bibinfo
  {author} {\bibfnamefont {M.}~\bibnamefont {Ghorbani-Asl}}, \bibinfo {author}
  {\bibfnamefont {L.}~\bibnamefont {Gammelgaard}}, \bibinfo {author}
  {\bibfnamefont {P.}~\bibnamefont {Bøggild}}, \bibinfo {author}
  {\bibfnamefont {K.~S.}\ \bibnamefont {Thygesen}}, \bibinfo {author}
  {\bibfnamefont {A.~V.}\ \bibnamefont {Krasheninnikov}}, \bibinfo {author}
  {\bibfnamefont {S.}~\bibnamefont {Xiao}}, \bibinfo {author} {\bibfnamefont
  {M.}~\bibnamefont {Wubs}},\ and\ \bibinfo {author} {\bibfnamefont
  {N.}~\bibnamefont {Stenger}},\ }\bibfield  {title} {\enquote {\bibinfo
  {title} {Controlled generation of luminescent centers in hexagonal boron
  nitride by irradiation engineering},}\ }\href
  {https://doi.org/10.1126/sciadv.abe7138} {\bibfield  {journal} {\bibinfo
  {journal} {Sci. Adv.}\ }\textbf {\bibinfo {volume} {7}},\ \bibinfo {pages}
  {eabe7138} (\bibinfo {year} {2021})}\BibitemShut {NoStop}%
\bibitem [{\citenamefont {Iv{\'a}dy}\ \emph {et~al.}(2020)\citenamefont
  {Iv{\'a}dy}, \citenamefont {Barcza}, \citenamefont {Thiering}, \citenamefont
  {Li}, \citenamefont {Hamdi}, \citenamefont {Chou}, \citenamefont {Legeza},\
  and\ \citenamefont {Gali}}]{Ivady2020}%
  \BibitemOpen
  \bibfield  {author} {\bibinfo {author} {\bibfnamefont {V.}~\bibnamefont
  {Iv{\'a}dy}}, \bibinfo {author} {\bibfnamefont {G.}~\bibnamefont {Barcza}},
  \bibinfo {author} {\bibfnamefont {G.}~\bibnamefont {Thiering}}, \bibinfo
  {author} {\bibfnamefont {S.}~\bibnamefont {Li}}, \bibinfo {author}
  {\bibfnamefont {H.}~\bibnamefont {Hamdi}}, \bibinfo {author} {\bibfnamefont
  {J.-P.}\ \bibnamefont {Chou}}, \bibinfo {author} {\bibfnamefont
  {{\"O}.}~\bibnamefont {Legeza}},\ and\ \bibinfo {author} {\bibfnamefont
  {A.}~\bibnamefont {Gali}},\ }\bibfield  {title} {\enquote {\bibinfo {title}
  {Ab initio theory of the negatively charged boron vacancy qubit in hexagonal
  boron nitride},}\ }\href {https://doi.org/10.1038/s41524-020-0305-x}
  {\bibfield  {journal} {\bibinfo  {journal} {NPJ Comput. Mater.}\ }\textbf
  {\bibinfo {volume} {6}},\ \bibinfo {pages} {41} (\bibinfo {year}
  {2020})}\BibitemShut {NoStop}%
\bibitem [{\citenamefont {Auburger}\ and\ \citenamefont
  {Gali}(2021)}]{Auburger2021}%
  \BibitemOpen
  \bibfield  {author} {\bibinfo {author} {\bibfnamefont {P.}~\bibnamefont
  {Auburger}}\ and\ \bibinfo {author} {\bibfnamefont {A.}~\bibnamefont
  {Gali}},\ }\bibfield  {title} {\enquote {\bibinfo {title} {Towards ab initio
  identification of paramagnetic substitutional carbon defects in hexagonal
  boron nitride acting as quantum bits},}\ }\href
  {https://doi.org/10.1103/PhysRevB.104.075410} {\bibfield  {journal} {\bibinfo
   {journal} {Phys. Rev. B}\ }\textbf {\bibinfo {volume} {104}},\ \bibinfo
  {pages} {075410} (\bibinfo {year} {2021})}\BibitemShut {NoStop}%
\bibitem [{\citenamefont {Mackoit-Sinkevičienė}\ \emph
  {et~al.}(2019)\citenamefont {Mackoit-Sinkevičienė}, \citenamefont
  {Maciaszek}, \citenamefont {de~Walle},\ and\ \citenamefont
  {Alkauskas}}]{Mackoit2019}%
  \BibitemOpen
  \bibfield  {author} {\bibinfo {author} {\bibfnamefont {M.}~\bibnamefont
  {Mackoit-Sinkevičienė}}, \bibinfo {author} {\bibfnamefont {M.}~\bibnamefont
  {Maciaszek}}, \bibinfo {author} {\bibfnamefont {C.~G.~V.}\ \bibnamefont
  {de~Walle}},\ and\ \bibinfo {author} {\bibfnamefont {A.}~\bibnamefont
  {Alkauskas}},\ }\bibfield  {title} {\enquote {\bibinfo {title} {Carbon dimer
  defect as a source of the 4.1 ev luminescence in hexagonal boron nitride},}\
  }\href {https://doi.org/10.1063/1.5124153} {\bibfield  {journal} {\bibinfo
  {journal} {Appl. Phys. Lett.}\ }\textbf {\bibinfo {volume} {115}},\ \bibinfo
  {pages} {212101} (\bibinfo {year} {2019})}\BibitemShut {NoStop}%
\bibitem [{\citenamefont {Gottscholl}\ \emph {et~al.}(2021)\citenamefont
  {Gottscholl}, \citenamefont {Diez}, \citenamefont {Soltamov}, \citenamefont
  {Kasper}, \citenamefont {Krauße}, \citenamefont {Sperlich}, \citenamefont
  {Kianinia}, \citenamefont {Bradac}, \citenamefont {Aharonovich},
  \citenamefont {Dyakonov},\ and\ \citenamefont {et~al.}}]{gottscholl}%
  \BibitemOpen
  \bibfield  {author} {\bibinfo {author} {\bibfnamefont {A.}~\bibnamefont
  {Gottscholl}}, \bibinfo {author} {\bibfnamefont {M.}~\bibnamefont {Diez}},
  \bibinfo {author} {\bibfnamefont {V.}~\bibnamefont {Soltamov}}, \bibinfo
  {author} {\bibfnamefont {C.}~\bibnamefont {Kasper}}, \bibinfo {author}
  {\bibfnamefont {D.}~\bibnamefont {Krauße}}, \bibinfo {author} {\bibfnamefont
  {A.}~\bibnamefont {Sperlich}}, \bibinfo {author} {\bibfnamefont
  {M.}~\bibnamefont {Kianinia}}, \bibinfo {author} {\bibfnamefont
  {C.}~\bibnamefont {Bradac}}, \bibinfo {author} {\bibfnamefont
  {I.}~\bibnamefont {Aharonovich}}, \bibinfo {author} {\bibfnamefont
  {V.}~\bibnamefont {Dyakonov}},\ and\ \bibinfo {author} {\bibnamefont
  {et~al.}},\ }\bibfield  {title} {\enquote {\bibinfo {title} {Spin defects in
  {hBN} as promising temperature, pressure and magnetic field quantum
  sensors},}\ }\href {https://doi.org/10.1038/s41467-021-24725-1} {\bibfield
  {journal} {\bibinfo  {journal} {Nat. Commun.}\ }\textbf {\bibinfo {volume}
  {12}},\ \bibinfo {pages} {4480} (\bibinfo {year} {2021})}\BibitemShut
  {NoStop}%
\bibitem [{\citenamefont {Stern}\ \emph {et~al.}(2022)\citenamefont {Stern},
  \citenamefont {Gu}, \citenamefont {Jarman}, \citenamefont {Eizagirre~Barker},
  \citenamefont {Mendelson}, \citenamefont {Chugh}, \citenamefont {Schott},
  \citenamefont {Tan}, \citenamefont {Sirringhaus}, \citenamefont
  {Aharonovich},\ and\ \citenamefont {et~al.}}]{stern}%
  \BibitemOpen
  \bibfield  {author} {\bibinfo {author} {\bibfnamefont {H.~L.}\ \bibnamefont
  {Stern}}, \bibinfo {author} {\bibfnamefont {Q.}~\bibnamefont {Gu}}, \bibinfo
  {author} {\bibfnamefont {J.}~\bibnamefont {Jarman}}, \bibinfo {author}
  {\bibfnamefont {S.}~\bibnamefont {Eizagirre~Barker}}, \bibinfo {author}
  {\bibfnamefont {N.}~\bibnamefont {Mendelson}}, \bibinfo {author}
  {\bibfnamefont {D.}~\bibnamefont {Chugh}}, \bibinfo {author} {\bibfnamefont
  {S.}~\bibnamefont {Schott}}, \bibinfo {author} {\bibfnamefont {H.~H.}\
  \bibnamefont {Tan}}, \bibinfo {author} {\bibfnamefont {H.}~\bibnamefont
  {Sirringhaus}}, \bibinfo {author} {\bibfnamefont {I.}~\bibnamefont
  {Aharonovich}},\ and\ \bibinfo {author} {\bibnamefont {et~al.}},\ }\bibfield
  {title} {\enquote {\bibinfo {title} {Room-temperature optically detected
  magnetic resonance of single defects in hexagonal boron nitride},}\ }\href
  {https://doi.org/10.1038/s41467-022-28169-z} {\bibfield  {journal} {\bibinfo
  {journal} {Nat. Commun.}\ }\textbf {\bibinfo {volume} {13}},\ \bibinfo
  {pages} {618} (\bibinfo {year} {2022})}\BibitemShut {NoStop}%
\bibitem [{\citenamefont {Liu}\ \emph {et~al.}(2019)\citenamefont {Liu},
  \citenamefont {Feng}, \citenamefont {Wang}, \citenamefont {Li},\ and\
  \citenamefont {Liu}}]{liu_feng_wang_li_liu_2019}%
  \BibitemOpen
  \bibfield  {author} {\bibinfo {author} {\bibfnamefont {G.-Q.}\ \bibnamefont
  {Liu}}, \bibinfo {author} {\bibfnamefont {X.}~\bibnamefont {Feng}}, \bibinfo
  {author} {\bibfnamefont {N.}~\bibnamefont {Wang}}, \bibinfo {author}
  {\bibfnamefont {Q.}~\bibnamefont {Li}},\ and\ \bibinfo {author}
  {\bibfnamefont {R.-B.}\ \bibnamefont {Liu}},\ }\bibfield  {title} {\enquote
  {\bibinfo {title} {Coherent quantum control of nitrogen-vacancy center spins
  near 1000 {Kelvin}},}\ }\href {https://doi.org/10.1038/s41467-019-09327-2}
  {\bibfield  {journal} {\bibinfo  {journal} {Nat. Commun.}\ }\textbf {\bibinfo
  {volume} {10}},\ \bibinfo {pages} {1344} (\bibinfo {year}
  {2019})}\BibitemShut {NoStop}%
\bibitem [{\citenamefont {Li}\ \emph {et~al.}(2019)\citenamefont {Li},
  \citenamefont {Xu}, \citenamefont {Mendelson}, \citenamefont {Kianinia},
  \citenamefont {Toth},\ and\ \citenamefont
  {Aharonovich}}]{LiXuMendelsonKianiniaTothAharonovich+2019+2049+2055}%
  \BibitemOpen
  \bibfield  {author} {\bibinfo {author} {\bibfnamefont {C.}~\bibnamefont
  {Li}}, \bibinfo {author} {\bibfnamefont {Z.-Q.}\ \bibnamefont {Xu}}, \bibinfo
  {author} {\bibfnamefont {N.}~\bibnamefont {Mendelson}}, \bibinfo {author}
  {\bibfnamefont {M.}~\bibnamefont {Kianinia}}, \bibinfo {author}
  {\bibfnamefont {M.}~\bibnamefont {Toth}},\ and\ \bibinfo {author}
  {\bibfnamefont {I.}~\bibnamefont {Aharonovich}},\ }\bibfield  {title}
  {\enquote {\bibinfo {title} {Purification of single-photon emission from hbn
  using post-processing treatments},}\ }\href
  {https://doi.org/doi:10.1515/nanoph-2019-0099} {\bibfield  {journal}
  {\bibinfo  {journal} {Nanophotonics}\ }\textbf {\bibinfo {volume} {8}},\
  \bibinfo {pages} {2049--2055} (\bibinfo {year} {2019})}\BibitemShut {NoStop}%
\bibitem [{\citenamefont {Roux}\ \emph {et~al.}(2022)\citenamefont {Roux},
  \citenamefont {Fournier}, \citenamefont {Watanabe}, \citenamefont
  {Taniguchi}, \citenamefont {Hermier}, \citenamefont {Barjon},\ and\
  \citenamefont {Delteil}}]{Roux2022}%
  \BibitemOpen
  \bibfield  {author} {\bibinfo {author} {\bibfnamefont {S.}~\bibnamefont
  {Roux}}, \bibinfo {author} {\bibfnamefont {C.}~\bibnamefont {Fournier}},
  \bibinfo {author} {\bibfnamefont {K.}~\bibnamefont {Watanabe}}, \bibinfo
  {author} {\bibfnamefont {T.}~\bibnamefont {Taniguchi}}, \bibinfo {author}
  {\bibfnamefont {J.-P.}\ \bibnamefont {Hermier}}, \bibinfo {author}
  {\bibfnamefont {J.}~\bibnamefont {Barjon}},\ and\ \bibinfo {author}
  {\bibfnamefont {A.}~\bibnamefont {Delteil}},\ }\bibfield  {title} {\enquote
  {\bibinfo {title} {Cathodoluminescence monitoring of quantum emitter
  activation in hexagonal boron nitride},}\ }\href
  {https://doi.org/10.1063/5.0126357} {\bibfield  {journal} {\bibinfo
  {journal} {Appl. Phys. Lett.}\ }\textbf {\bibinfo {volume} {121}},\ \bibinfo
  {pages} {184002} (\bibinfo {year} {2022})}\BibitemShut {NoStop}%
\bibitem [{\citenamefont {Gr\"unwald}(2019)}]{Gr_nwald_2019}%
  \BibitemOpen
  \bibfield  {author} {\bibinfo {author} {\bibfnamefont {P.}~\bibnamefont
  {Gr\"unwald}},\ }\bibfield  {title} {\enquote {\bibinfo {title} {Effective
  second-order correlation function and single-photon detection},}\ }\href
  {https://doi.org/10.1088/1367-2630/ab3ae0} {\bibfield  {journal} {\bibinfo
  {journal} {New J. Phys.}\ }\textbf {\bibinfo {volume} {21}},\ \bibinfo
  {pages} {093003} (\bibinfo {year} {2019})}\BibitemShut {NoStop}%
\bibitem [{\citenamefont {Fishman}\ \emph {et~al.}(2023)\citenamefont
  {Fishman}, \citenamefont {Patel}, \citenamefont {Hopper}, \citenamefont
  {Huang},\ and\ \citenamefont {Bassett}}]{arXiv:2111.01252}%
  \BibitemOpen
  \bibfield  {author} {\bibinfo {author} {\bibfnamefont {R.~E.}\ \bibnamefont
  {Fishman}}, \bibinfo {author} {\bibfnamefont {R.~N.}\ \bibnamefont {Patel}},
  \bibinfo {author} {\bibfnamefont {D.~A.}\ \bibnamefont {Hopper}}, \bibinfo
  {author} {\bibfnamefont {T.-Y.}\ \bibnamefont {Huang}},\ and\ \bibinfo
  {author} {\bibfnamefont {L.~C.}\ \bibnamefont {Bassett}},\ }\bibfield
  {title} {\enquote {\bibinfo {title} {Photon-emission-correlation spectroscopy
  as an analytical tool for solid-state quantum defects},}\ }\href
  {https://doi.org/10.1103/PRXQuantum.4.010202} {\bibfield  {journal} {\bibinfo
   {journal} {PRX Quantum}\ }\textbf {\bibinfo {volume} {4}},\ \bibinfo {pages}
  {010202} (\bibinfo {year} {2023})}\BibitemShut {NoStop}%
\bibitem [{\citenamefont {Vogl}\ \emph
  {et~al.}(2019{\natexlab{c}})\citenamefont {Vogl}, \citenamefont {Lecamwasam},
  \citenamefont {Buchler}, \citenamefont {Lu},\ and\ \citenamefont
  {Lam}}]{vogl_lecamwasam_buchler_lu_lam_2019}%
  \BibitemOpen
  \bibfield  {author} {\bibinfo {author} {\bibfnamefont {T.}~\bibnamefont
  {Vogl}}, \bibinfo {author} {\bibfnamefont {R.}~\bibnamefont {Lecamwasam}},
  \bibinfo {author} {\bibfnamefont {B.~C.}\ \bibnamefont {Buchler}}, \bibinfo
  {author} {\bibfnamefont {Y.}~\bibnamefont {Lu}},\ and\ \bibinfo {author}
  {\bibfnamefont {P.~K.}\ \bibnamefont {Lam}},\ }\bibfield  {title} {\enquote
  {\bibinfo {title} {Compact cavity-enhanced single-photon generation with
  hexagonal boron nitride},}\ }\href
  {https://doi.org/10.1021/acsphotonics.9b00314} {\bibfield  {journal}
  {\bibinfo  {journal} {ACS Photonics}\ }\textbf {\bibinfo {volume} {6}},\
  \bibinfo {pages} {1955–1962} (\bibinfo {year}
  {2019}{\natexlab{c}})}\BibitemShut {NoStop}%
\bibitem [{\citenamefont {Boll}\ \emph {et~al.}(2020)\citenamefont {Boll},
  \citenamefont {Radko}, \citenamefont {Huck},\ and\ \citenamefont
  {Andersen}}]{Boll:20}%
  \BibitemOpen
  \bibfield  {author} {\bibinfo {author} {\bibfnamefont {M.~K.}\ \bibnamefont
  {Boll}}, \bibinfo {author} {\bibfnamefont {I.~P.}\ \bibnamefont {Radko}},
  \bibinfo {author} {\bibfnamefont {A.}~\bibnamefont {Huck}},\ and\ \bibinfo
  {author} {\bibfnamefont {U.~L.}\ \bibnamefont {Andersen}},\ }\bibfield
  {title} {\enquote {\bibinfo {title} {Photophysics of quantum emitters in
  hexagonal boron-nitride nano-flakes},}\ }\href
  {https://doi.org/10.1364/OE.386629} {\bibfield  {journal} {\bibinfo
  {journal} {Opt. Express}\ }\textbf {\bibinfo {volume} {28}},\ \bibinfo
  {pages} {7475--7487} (\bibinfo {year} {2020})}\BibitemShut {NoStop}%
\bibitem [{\citenamefont {Stern}\ \emph {et~al.}(2019)\citenamefont {Stern},
  \citenamefont {Wang}, \citenamefont {Fan}, \citenamefont {Mizuta},
  \citenamefont {Stewart}, \citenamefont {Needham}, \citenamefont {Roberts},
  \citenamefont {Wai}, \citenamefont {Ginsberg}, \citenamefont {Klenerman},
  \citenamefont {Hofmann},\ and\ \citenamefont
  {Lee}}]{doi:10.1021/acsnano.9b00274}%
  \BibitemOpen
  \bibfield  {author} {\bibinfo {author} {\bibfnamefont {H.~L.}\ \bibnamefont
  {Stern}}, \bibinfo {author} {\bibfnamefont {R.}~\bibnamefont {Wang}},
  \bibinfo {author} {\bibfnamefont {Y.}~\bibnamefont {Fan}}, \bibinfo {author}
  {\bibfnamefont {R.}~\bibnamefont {Mizuta}}, \bibinfo {author} {\bibfnamefont
  {J.~C.}\ \bibnamefont {Stewart}}, \bibinfo {author} {\bibfnamefont {L.-M.}\
  \bibnamefont {Needham}}, \bibinfo {author} {\bibfnamefont {T.~D.}\
  \bibnamefont {Roberts}}, \bibinfo {author} {\bibfnamefont {R.}~\bibnamefont
  {Wai}}, \bibinfo {author} {\bibfnamefont {N.~S.}\ \bibnamefont {Ginsberg}},
  \bibinfo {author} {\bibfnamefont {D.}~\bibnamefont {Klenerman}}, \bibinfo
  {author} {\bibfnamefont {S.}~\bibnamefont {Hofmann}},\ and\ \bibinfo {author}
  {\bibfnamefont {S.~F.}\ \bibnamefont {Lee}},\ }\bibfield  {title} {\enquote
  {\bibinfo {title} {Spectrally resolved photodynamics of individual emitters
  in large-area monolayers of hexagonal boron nitride},}\ }\href
  {https://doi.org/10.1021/acsnano.9b00274} {\bibfield  {journal} {\bibinfo
  {journal} {ACS Nano}\ }\textbf {\bibinfo {volume} {13}},\ \bibinfo {pages}
  {4538--4547} (\bibinfo {year} {2019})}\BibitemShut {NoStop}%
\bibitem [{\citenamefont {Mu}\ \emph {et~al.}(2022)\citenamefont {Mu},
  \citenamefont {Cai}, \citenamefont {Chen}, \citenamefont {Kenny},
  \citenamefont {Jiang}, \citenamefont {Ru}, \citenamefont {Lyu}, \citenamefont
  {Koh}, \citenamefont {Liu}, \citenamefont {Aharonovich},\ and\ \citenamefont
  {Gao}}]{Mu2022}%
  \BibitemOpen
  \bibfield  {author} {\bibinfo {author} {\bibfnamefont {Z.}~\bibnamefont
  {Mu}}, \bibinfo {author} {\bibfnamefont {H.}~\bibnamefont {Cai}}, \bibinfo
  {author} {\bibfnamefont {D.}~\bibnamefont {Chen}}, \bibinfo {author}
  {\bibfnamefont {J.}~\bibnamefont {Kenny}}, \bibinfo {author} {\bibfnamefont
  {Z.}~\bibnamefont {Jiang}}, \bibinfo {author} {\bibfnamefont
  {S.}~\bibnamefont {Ru}}, \bibinfo {author} {\bibfnamefont {X.}~\bibnamefont
  {Lyu}}, \bibinfo {author} {\bibfnamefont {T.~S.}\ \bibnamefont {Koh}},
  \bibinfo {author} {\bibfnamefont {X.}~\bibnamefont {Liu}}, \bibinfo {author}
  {\bibfnamefont {I.}~\bibnamefont {Aharonovich}},\ and\ \bibinfo {author}
  {\bibfnamefont {W.}~\bibnamefont {Gao}},\ }\bibfield  {title} {\enquote
  {\bibinfo {title} {Excited-state optically detected magnetic resonance of
  spin defects in hexagonal boron nitride},}\ }\href
  {https://doi.org/10.1103/PhysRevLett.128.216402} {\bibfield  {journal}
  {\bibinfo  {journal} {Phys. Rev. Lett.}\ }\textbf {\bibinfo {volume} {128}},\
  \bibinfo {pages} {216402} (\bibinfo {year} {2022})}\BibitemShut {NoStop}%
\bibitem [{\citenamefont {Kresse}\ and\ \citenamefont
  {Furthmüller}(1996)}]{vasp1}%
  \BibitemOpen
  \bibfield  {author} {\bibinfo {author} {\bibfnamefont {G.}~\bibnamefont
  {Kresse}}\ and\ \bibinfo {author} {\bibfnamefont {J.}~\bibnamefont
  {Furthmüller}},\ }\bibfield  {title} {\enquote {\bibinfo {title} {Efficiency
  of ab-initio total energy calculations for metals and semiconductors using a
  plane-wave basis set},}\ }\href
  {https://doi.org/https://doi.org/10.1016/0927-0256(96)00008-0} {\bibfield
  {journal} {\bibinfo  {journal} {Comput. Mater. Sci.}\ }\textbf {\bibinfo
  {volume} {6}},\ \bibinfo {pages} {15 -- 50} (\bibinfo {year}
  {1996})}\BibitemShut {NoStop}%
\bibitem [{\citenamefont {Kresse}\ and\ \citenamefont
  {Furthm\"uller}(1996)}]{vasp2}%
  \BibitemOpen
  \bibfield  {author} {\bibinfo {author} {\bibfnamefont {G.}~\bibnamefont
  {Kresse}}\ and\ \bibinfo {author} {\bibfnamefont {J.}~\bibnamefont
  {Furthm\"uller}},\ }\bibfield  {title} {\enquote {\bibinfo {title} {Efficient
  iterative schemes for ab initio total-energy calculations using a plane-wave
  basis set},}\ }\href
  {https://doi.org/https://doi.org/10.1103/PhysRevB.54.11169} {\bibfield
  {journal} {\bibinfo  {journal} {Phys. Rev. B}\ }\textbf {\bibinfo {volume}
  {54}},\ \bibinfo {pages} {11169--11186} (\bibinfo {year} {1996})}\BibitemShut
  {NoStop}%
\bibitem [{\citenamefont {Jara}\ \emph {et~al.}(2020)\citenamefont {Jara},
  \citenamefont {Rauch}, \citenamefont {Botti}, \citenamefont {Marques},
  \citenamefont {Norambuena}, \citenamefont {Coto}, \citenamefont {Maze},\ and\
  \citenamefont {Munoz}}]{Jara2020}%
  \BibitemOpen
  \bibfield  {author} {\bibinfo {author} {\bibfnamefont {C.}~\bibnamefont
  {Jara}}, \bibinfo {author} {\bibfnamefont {T.}~\bibnamefont {Rauch}},
  \bibinfo {author} {\bibfnamefont {S.}~\bibnamefont {Botti}}, \bibinfo
  {author} {\bibfnamefont {M.~A.}\ \bibnamefont {Marques}}, \bibinfo {author}
  {\bibfnamefont {A.}~\bibnamefont {Norambuena}}, \bibinfo {author}
  {\bibfnamefont {R.}~\bibnamefont {Coto}}, \bibinfo {author} {\bibfnamefont
  {J.~R.}\ \bibnamefont {Maze}},\ and\ \bibinfo {author} {\bibfnamefont
  {F.}~\bibnamefont {Munoz}},\ }\bibfield  {title} {\enquote {\bibinfo {title}
  {First-principles identification of single-photon emitters based on carbon
  clusters in hexagonal boron nitride},}\ }\href@noop {} {\bibfield  {journal}
  {\bibinfo  {journal} {J. Phys. Chem. A}\ } (\bibinfo {year}
  {2020})}\BibitemShut {NoStop}%
\bibitem [{\citenamefont {Li}, \citenamefont {Smart},\ and\ \citenamefont
  {Ping}(2022)}]{Li2022}%
  \BibitemOpen
  \bibfield  {author} {\bibinfo {author} {\bibfnamefont {K.}~\bibnamefont
  {Li}}, \bibinfo {author} {\bibfnamefont {T.~J.}\ \bibnamefont {Smart}},\ and\
  \bibinfo {author} {\bibfnamefont {Y.}~\bibnamefont {Ping}},\ }\bibfield
  {title} {\enquote {\bibinfo {title} {Carbon trimer as a 2 ev single-photon
  emitter candidate in hexagonal boron nitride: A first-principles study},}\
  }\href {https://doi.org/10.1103/PhysRevMaterials.6.L042201} {\bibfield
  {journal} {\bibinfo  {journal} {Phys. Rev. Mater.}\ }\textbf {\bibinfo
  {volume} {6}} (\bibinfo {year} {2022}),\
  10.1103/PhysRevMaterials.6.L042201}\BibitemShut {NoStop}%
\bibitem [{\citenamefont {Vuong}\ \emph {et~al.}(2016)\citenamefont {Vuong},
  \citenamefont {Cassabois}, \citenamefont {Valvin}, \citenamefont {Ouerghi},
  \citenamefont {Chassagneux}, \citenamefont {Voisin},\ and\ \citenamefont
  {Gil}}]{Vuong2016}%
  \BibitemOpen
  \bibfield  {author} {\bibinfo {author} {\bibfnamefont {T.~Q.~P.}\
  \bibnamefont {Vuong}}, \bibinfo {author} {\bibfnamefont {G.}~\bibnamefont
  {Cassabois}}, \bibinfo {author} {\bibfnamefont {P.}~\bibnamefont {Valvin}},
  \bibinfo {author} {\bibfnamefont {A.}~\bibnamefont {Ouerghi}}, \bibinfo
  {author} {\bibfnamefont {Y.}~\bibnamefont {Chassagneux}}, \bibinfo {author}
  {\bibfnamefont {C.}~\bibnamefont {Voisin}},\ and\ \bibinfo {author}
  {\bibfnamefont {B.}~\bibnamefont {Gil}},\ }\bibfield  {title} {\enquote
  {\bibinfo {title} {Phonon-photon mapping in a color center in hexagonal boron
  nitride},}\ }\href {https://doi.org/10.1103/PhysRevLett.117.097402}
  {\bibfield  {journal} {\bibinfo  {journal} {Phys. Rev. Lett.}\ }\textbf
  {\bibinfo {volume} {117}},\ \bibinfo {pages} {097402} (\bibinfo {year}
  {2016})}\BibitemShut {NoStop}%
\bibitem [{\citenamefont {Tan}\ \emph {et~al.}(2022)\citenamefont {Tan},
  \citenamefont {Lai}, \citenamefont {Liu}, \citenamefont {Guo}, \citenamefont
  {Xue}, \citenamefont {Dou}, \citenamefont {Sun}, \citenamefont {Deng},
  \citenamefont {Tan}, \citenamefont {Aharonovich}, \citenamefont {Gao},\ and\
  \citenamefont {Zhang}}]{doi:10.1021/acs.nanolett.1c04647}%
  \BibitemOpen
  \bibfield  {author} {\bibinfo {author} {\bibfnamefont {Q.}~\bibnamefont
  {Tan}}, \bibinfo {author} {\bibfnamefont {J.-M.}\ \bibnamefont {Lai}},
  \bibinfo {author} {\bibfnamefont {X.-L.}\ \bibnamefont {Liu}}, \bibinfo
  {author} {\bibfnamefont {D.}~\bibnamefont {Guo}}, \bibinfo {author}
  {\bibfnamefont {Y.}~\bibnamefont {Xue}}, \bibinfo {author} {\bibfnamefont
  {X.}~\bibnamefont {Dou}}, \bibinfo {author} {\bibfnamefont {B.-Q.}\
  \bibnamefont {Sun}}, \bibinfo {author} {\bibfnamefont {H.-X.}\ \bibnamefont
  {Deng}}, \bibinfo {author} {\bibfnamefont {P.-H.}\ \bibnamefont {Tan}},
  \bibinfo {author} {\bibfnamefont {I.}~\bibnamefont {Aharonovich}}, \bibinfo
  {author} {\bibfnamefont {W.}~\bibnamefont {Gao}},\ and\ \bibinfo {author}
  {\bibfnamefont {J.}~\bibnamefont {Zhang}},\ }\bibfield  {title} {\enquote
  {\bibinfo {title} {Donor–acceptor pair quantum emitters in hexagonal boron
  nitride},}\ }\href@noop {} {\bibfield  {journal} {\bibinfo  {journal} {Nano
  Lett.}\ }\textbf {\bibinfo {volume} {22}},\ \bibinfo {pages} {1331--1337}
  (\bibinfo {year} {2022})}\BibitemShut {NoStop}%
\end{thebibliography}%

\end{document}